\documentclass[12pt]{article}
\usepackage{amssymb,epsfig}

\newcommand{\non}{\nonumber \\}

\newcommand{\R}[1]{(\ref{eq:#1})}

\newcommand{\ket}[1]{|#1 \rangle}

\newcommand{\sign}[1]{\mbox{sign}(#1)}

\def\a{\alpha}
\def\b{\beta}
\def\c{\gamma}
\def\d{\delta}
\def\e{\epsilon}

\def\h{\eta}
\def\i{{\rm i}}

\def\l{\lambda}
\def\m{\mu}

\def\r{\rho}
\def\s{\sigma}
\def\t{\tau}

\def\C{\Gamma}

\def\L{\Lambda}

\def\T{\Xi}

\def\cA{{\cal A}}  \def\cC{{\cal C}}
  \def\cF{{\cal F}}
\def\cG{{\cal G}}  
 \def\cK{{\cal K}} 
\def\cM{{\cal M}} \def\cN{{\cal N}} \def\cO{{\cal O}}
\def\cP{{\cal P}} \def\cQ{{\cal Q}} 
\def\cS{{\cal S}} \def\cT{{\cal T}} 
 \def\cW{{\cal W}} \def\cX{{\cal X}}
 \def\cZ{{\cal Z}}

\def\rar{\rightarrow}

\def\hlf{\frac{1}{2}}

\def\ZZ{\mathbb{Z}}

\def\beq{\begin{equation}}
\def\eeq{\end{equation}}

\def\beqx{\begin{displaymath}}
\def\eeqx{\end{displaymath}}

\def\bea{\begin{eqnarray}}
\def\eea{\end{eqnarray}}

\def\mod{\;{\rm mod}\;\;}
\def\i{{\rm i}}

\def\bg{\bar{\mathfrak{g}}}
\def\bL{{\bar{\Lambda}}}

\def\tW{{\tilde W}}

\def\[{\left [}
\def\]{\right ]}
\def\({\left(}
\def\){\right)}
\def\tG{\tilde{G}}

\begin{document}
\begin{flushright}
\parbox[t]{2in}{NIKHEF-Th-01-016\\
    CU-TP-1040\\
  hep-th/0110267}
\end{flushright}

\vspace*{0.2in}

\begin{center}
{\Large \bf Geometry of WZW Orientifolds
}

\vspace*{0.7in}
{L.R. Huiszoon${}^{1}$,~K. Schalm${}^{1,2}$~
and A.N. Schellekens${}^{1}$}\\[.5in]
{\em ${}^1$ NIKHEF Theory Group \\
P.O. Box 41882\\
1009 DB Amsterdam}\\[0.25in]
{\em ${}^2$Department of Physics~\footnote[3]{Address
    since
September 1st.} \\
Columbia University\\
New York, NY 10027}\\[1in]
\end{center}

\begin{center}
{\bf
Abstract}
\end{center}

We analyze unoriented Wess-Zumino-Witten models from a geometrical
point of view.
We show that
the geometric interpretation of simple current crosscap states is as
centre
orientifold planes localized on conjugacy classes
of the group manifold. We determine the locations and dimensions of
these planes for arbitrary simply-connected groups and orbifolds
thereof. The dimensions of the O-planes turn out to be given by the
dimensions
of symmetric coset manifolds based on regular embeddings.
Furthermore, we give a geometrical interpretation of boundary
conjugation in
open unoriented WZW models; it yields D-branes together with their
 images under the orientifold projection.
To find the agreement between O-planes and
crosscap states, we find explicit answers for lattice extensions of Gaussian
sums. These results allow us to express the modular $P$-matrix, which is
directly related to the crosscap coefficient, in terms of
characters of the horizontal subgroup of the affine Lie algebra. A
corollary of this relation is that there exists a formal linear relation
between the modular $P$- and the modular $S$-matrix.

\newpage
\section{Introduction}

There has been much progress in the study of interacting conformal field
theories on open unoriented surfaces~\cite{cardy,ishibashi,sagnotti,unor}.
Compared with the closed case, this
requires the introduction of two
new objects, boundary and crosscap states,
which act as sources for closed strings.
Starting with~\cite{douglas}, there has been much interest in the geometrical  
interpretation of
the first of these two: the boundary states. By extension of the toroidal
results, they are believed to give a
stringy description of D-branes~\cite{PolRR} in curved
backgrounds. Owing
to the presence of
localized Yang-Mills fields on their surface, their natural geometric
interpretation is as vector-bundles wrapping closed submanifolds or
more precisely coherent
 sheaves. And owing to the geometric interpretation the abstract CFT
 results may be extrapolated away from the special rational point
to other areas of the moduli
 space of the theory.

In many realistic string compactifications however, the presence of boundary
states leads to tadpoles, and we are forced to include crosscap
states in CFT.  In geometric language one needs to absorb the excess
charges carried by the D-branes by the introduction of (some superposition of)
orientifold planes~\cite{Oplanes}. Outside of
  toroidal or orbifold compactifications, little is known about the
geometry of orientifolds (see \cite{ghoshal} for a first
attempt to study O-planes in a curved geometry). It is not clear
whether orientifolds also should be interpreted as independent
objects wrapping closed submanifolds. From a CFT point of view
it is {\em a priori} not
even clear that crosscap states yield properly
localized charge distributions which may be interpreted as
(lower-dimensional) orientifold planes.
Known examples of toroidal or orbifold models are limited to
special points in the stringy region of the geometric moduli-space
where the target space is flat except for some
singular points. Indeed from a perturbative string-theory point of view
orientifolds seem little more than a bookkeeping device for excess
curvature or charge, because they have no inherent moduli and are
strictly rigid objects. On the other
hand, non-perturbative
studies do indicate
that they may possess some properties which allow them to be seen as
bound states of independent higher-dimensional objects in
string theory. An example hereof is the splitting of O7-planes into
D7-branes under splitting of the singularities of F-theory
compactifications on K3 \cite{sen}.

Here we wish to make a first step towards determining the geometrical
characteristics of crosscap states. Understanding these would allow us
to extend unoriented type I compactifications to regions away from
rational points in the moduli space, and make contact with large
volume geometry. We have to be careful,
for it has not yet been established whether this is a truly sensible
question to ask. The orientation reversal projection, characteristic of the
crosscap state, may project out some blow-up modes which are
necessary to reach the large volume point of the moduli space. Hence the
target-space geometry, or more precisely its moduli space, of unoriented
models
is not directly equivalent to that of its oriented parent. In turn this means
that we should also reconsider the geometry of boundary states in
unoriented models, as it may need reinterpretation.

In this article, we will start to
address these questions by studying orientifold
planes on group manifolds. Strings on group manifolds are described by
Wess-Zumino-Witten (WZW)
models, which are a well-understood class of interacting CFT's. In
particular the crosscap states which may arise in WZW-models are
known. On the other hand,
to consider the group manifold of a WZW-model
as target spaces does not always agree with the conventional
interpretation of the target space of string theories.
Most importantly, in standard string scenarios we
think of the string moving in a target space with dimension equal to
the central charge of the bosonic CFT. With WZW models, on the other
hand, one thinks of the string living on the group manifold with
dimension that of the group itself.
The word ``geometry'' depends on the setting in which it is discussed.

We will proceed with this caveat in mind. In section
2 we will use simple geometrical (group theoretical) techniques in
the Lagrangian description of WZW models to classify the
possible WZW orientifold planes by their locations and dimensions. For every
element of the centre of the group, we find a configuration of O-planes that
are localized on conjugacy classes. In section 3, we review the
construction of abstract unoriented CFT's in general and WZW-models in
particular. We show that in order to determine the characteristics of WZW
crosscap states, we need an expression for the pseudo-modular $P$-matrix in
terms of
characters of the horizontal subalgebra. This specific mathematical
problem we answer in section 4, whereafter in section 5 we reach the
objective of this article. We
demonstrate that there is indeed a one-to-one correspondence between
the so-called simple current crosscap states~\cite{klein} of
Wess-Zumino-Witten models and the geometrical O-planes. We also show that the  
crosscap states of~\cite{FOE} describe configurations of O-planes on
non-simply connected group manifolds (orbifolds). In the final
section, section 6,  we
describe the geometry of branes in {\em un}oriented WZW models, thereby
completing the earlier study of branes on group
manifolds~\cite{alSc,brWZW,stanciu,Malda-etal}.

The important calculation of section 4 that expresses the $P$-matrix of a WZW-model
in terms of characters of the horizontal subalgebra is the heart of
this paper. There are at least two reasons for us to
surmise that it may have consequences beyond the
immediate question posed here. In deriving the relation
between the $P$-matrix and characters, one
has to perform mathematically
interesting `quadratic Gaussian sums' on the weight lattices of
the horizontal algebra. These are extensions of
 one-dimensional Gaussian sums and $d$-Cartesian dimensional
 Kloosterman sums, which play a role
in analytic number
theory. Physically interesting is the fact that the relation with
characters of $G$ implies a
linear relation between the modular $P$- and the modular $S$-matrix for WZW-models,
albeit an analytic continuation of the latter, which we define
in section (4.2).
The
construction of open-string theories relies heavily on the properties
of both, and if the existence of such a linear relation were to hold
for all abstract CFT's, this could have important consequences, for instance
for tadpole cancellation.

Concerning the notation: a tilde is used for the rightmoving sector,
but is often omitted for the sake of clarity. The affine algebra will
be denoted by $\mathfrak{g}$ and the horizontal subalgebra by
$\bar{\mathfrak{g}}$. By extension all affine quantities are unbarred,
for
instance weights $\L$ and characters $\chi$, and all horizontal
quantities are barred. However, when the formulas are unambiguous, we
will
often drop the bars. Lie group characters are denoted by $\cX$.

Two final remarks: One, to prevent an overcluttered notation we have
normalized the length squared of the highest root $\Psi$ for every
group to
two: $(\Psi,\Psi)=2$. Two,
in trying to keep the presentation as simple as possible, we
will assume that the group centres are cyclic.
At the end of some sections, we will comment on the generalization to
noncyclic
 centres,
that occur for $D_r,~r$ even, or tensor products of WZW models.

While we were applying the final touches, a paper \cite{brunner}
appeared
which also finds some of the results discussed here.

\section{Semiclassical analysis} \label{sec-semi}

WZW models, as do free CFT's, have the advantage that there also exists
a Lagrangian
description of the theory.
String theory on a group manifold $G$ can either be described by the
abstract CFT built on an affine extension $\mathfrak{g}$ of the algebra
$\bar{\mathfrak{g}}$ of $G$ or by the
Wess-Zumino-Witten action \cite{witten,goddard}
\begin{equation}
\label{action}
    S = -\frac{k}{16\pi} \int_{\Sigma} d^2x {\rm Tr}
    (g^{-1}\partial_{\mu} g)^2 + \frac{k}{24\pi} \int_B d^3y {\rm Tr}
    (g^{-1}dg)^3 \;\;\; .
\end{equation}
Here $g$ is a field that takes values in the group $G$
and lives on the string worldsheet $\Sigma$. The second term is
  topological; $B$ is a three-manifold whose boundary is $\Sigma$, and the level  
$k$ has to be integer. The Lagrangian description emphasizes the
geometry and it is therefore the natural place to explore the
geometrical consequences of orientation reversal.

The WZW
theory possesses the chiral currents~\cite{witten}
\begin{equation}
    W = g^{-1} \partial g \;\;,\;\; \tW = - \tilde{\partial} gg^{-1}~,
\end{equation}
whose modes span two commuting copies of the affine Lie algebra $\mathfrak{g}$. 
To construct an unoriented theory, consider the
involutions
\begin{equation} \label{eq:Ogroup}
    \Omega_n := \Omega \cdot R_n \;\;\;\; ,
\end{equation}
The worldsheet parity transformation $\Omega$ interchanges the
chiral coordinates and $R_n$ acts on the coordinates as
\begin{equation} \label{eq:gdef}
    R_n : g \rightarrow \c^n g^{-1} \;\;\;\; , \;\;\;\; n=0,...,N-1
\end{equation}
where $\c$ generates the $\ZZ_N$ centre of $G$. $\Omega_n$
interchanges the chiral currents and is therefore a global
invariance of this theory. When we gauge this symmetry, we obtain
orientifold fixed planes at those points of the target
space $G$ that are fixed under $\Omega_n$~\cite{Oplanes}.
Note that $\Omega$ by itself is
not a symmetry. This tells us right away that WZW models do not possess space-time
filling
orientifold-planes, analogous to O9-planes in superstring theory. Instead,
we only get lower dimensional planes at the points $g\in G$ for which
\begin{equation} \label{eq:loc}
    g = \c^n g^{-1} \;\;\;\; .
\end{equation}
Note that, if $g$ solves~(\ref{eq:loc}), so does every point in the set
\begin{equation} \label{eq:conju}
    C(g) = \{hgh^{-1}|h\in G\} \;\;\;\; ,
\end{equation}
so our planes are located at conjugacy classes of $G$.

To solve~(\ref{eq:loc}) explicitly, recall that
every conjugacy class contains at least one element of a maximal torus
$T$ of $G$.
So choose a Chevalley basis of the rank $r$
algebra $\bg$ and let $H^i,i=1,...,r$ be a basis for the Cartan
subalgebra. In this basis their action on the weights returns the
integral Dynkin labels.
Then an element of the corresponding maximal torus can be written as
 $g_{t} = \exp{(2\pi\i (t,H))} := \exp{(2\pi\i \sum_{i=1}^r t_i
  H^i)}$. In this
basis, the centre elements $\c^n$ are
given by
\beq \label{eq:cc}
    \c^n = \exp\{2\pi\i {n \over N} \sum_i c_i H^i\}~,
\eeq
where $\cC := \sum_i c_i H^i$ is tabulated in table \ref{tab:O-planes}. The
eigenvalue of $\cC$ modulo $N$ is known as the conjugacy
class of a representation\footnote{Not to be confused with
the conjugacy classes for groups used above.} of the algebra
$\bar{\mathfrak{g}}$. It is constant modulo $N$ within a
representation, which means that $\cC$ commutes modulo $N$ with all
other generators in that representation. This shows that $\c^n$ indeed
belongs to the centre (and, by turning the argument around, that the
elements of the centre of $G$ are in one-to-one correspondence with the
conjugacy classes of $\bar{\mathfrak{g}}$.)

It is now easy to find the solutions to~(\ref{eq:loc}). An element
$g_t$ is a fixed point of
$R_{n}$ when  for all $i$
\beq \label{eq:sol}
    2t_i = \frac{n c_i}{N} \mod 1 \;\;\; \rightarrow \;\;\; t_i =
    \frac{nc_i}{2N} - \frac{u_i}{2} \mod 1\;\;\;\; .
\eeq
Here $u_i$ are vectors in a basis dual to the Dynkin basis,
whose entries are either zero or one modulo even
integers. We can phrase this in a basis-independent way, using
that the dual of the weight space $L_w$ is the coroot
lattice $L^\vee$. Eq. (\ref{eq:sol}) is a vector equation on a basis
of $r$ simple
coroots $\tilde{\a}^i,~i=1,\dots,r$;
\begin{equation}
  \label{eq:1}
  2 t = \frac{nc}{N} \mod L^\vee \;\;\; \rightarrow \;\;\; t =
    \frac{nc}{2N} - \frac{u}{2} \;\;\;\; {\rm with}~~ ~u \in L^\vee \mod
     2L^\vee~~,
\end{equation}
with $t=\sum_i t_i\tilde{\a}^i$, etc.
The minus sign in front of $u$ is chosen for later
purposes (to make contact with~\R{g}). So we have found $2^r$
solutions
$g_{n,u}$ to~(\ref{eq:loc}). In general, however, not
all of them lie on different conjugacy classes. A quite non-trivial
theorem states that when two elements of the maximal torus are
conjugate, they are related by the action of the Weyl group
$W$~\cite{books}. Now we can count how many and what type of O-planes we get.

Consider first the projection with $n=0$ and define
\beq \label{eq:nis0}
    g_u := g_{0,u} = e^{-\pi\i (u,H)}~.
\eeq
Then, according to this  theorem, $g_u$ and $g_{u'}$ lie on the same
conjugacy class when $u$ and $u'$  belong to the same {\em Weyl orbit}, i.e.
when there is a $w\in W$ such
that
\beq \label{eq:weylorb}
    u' = w(u) \;\;\;\; \mod 2L^\vee.
\eeq
So the number of O-planes equals the number of Weyl orbits of $\bg$. We
determined these numbers empirically and displayed them in table
\ref{tab:O-planes}.

\begin{table}[t]
\begin{center}
\small
    \begin{tabular}{|l||l|l|l|} \hline
$\mathfrak{g}$ & Current $J$       & $\cC$             & $\#$ O-planes
\\ \hline \hline
$A_r$, $r$ even & --     & 0      & $1 + \frac{r}{2}$ \hfill
\\ \hline
$A_r$, $r$ even & $(k,0,\ldots,0)$     & $\sum_{i=1}^r iH^i$       & $1 +
\frac{r}{2} $\hfill $(*)$
\\ \hline
$A_r$, $r$ odd & --     & 0       & $1 + \frac{(r+1)}{2}$\hfill
\\ \hline
$A_r$, $r$ odd & $(k,0,\ldots,0)$     & $\sum_{i=1}^r iH^i$       & $1 +
\frac{(r-1)}{2}$\hfill
\\ \hline
$B_r$, $r$ even & --     & 0             & $2 + \frac{r}{2}$ \hfill  \\ \hline
$B_r$, $r$ even & $(k,0,\ldots,0)$     & $H^r$             &
$\frac{r}{2}$\hfill   \\ \hline
$B_r$, $r$ odd & --     & 0             & $1 + \frac{(r+1)}{2}$ \hfill  \\ \hline
$B_r$, $r$ odd  & $(k,0,\ldots,0)$    & $H^r$             &
$\frac{(r+1)}{2}$  \hfill \\ \hline
$C_r$ & --     & 0 & $1 + r$\hfill  \\ \hline
$C_r$ & $(0,\ldots,0,k)$     & $\sum_{i=1}^{[(r+1)/2]} H^{2i-1}$ & $1$\hfill
 \\ \hline
$D_r$, $r$ even & --  & 0
&  $3+\frac{r}{2}$\hfill \\ \hline
$D_r$, $r$ even & $(0,\ldots,0,k)$  & $2\sum_{i=1}^{r/2-1}
H^{2i-1}+(2\!-\!r)H^{r-1}-rH^r$
&  $2$\hfill \\ \hline
$D_r$, $r$ even & $(k,0,\ldots,0) $  & $2H^{r-1}+2H^r$
&  $\frac{r}{2}$\hfill \\ \hline
$D_r$, $r$ even & $(0,\ldots,0,k,0)$  & $2\sum_{i=1}^{r/2-1}
H^{2i-1}-rH^{r-1}+(2\!-\!r)H^r$
&  $2$\hfill \\ \hline
$D_r$, $r$ odd & --  & 0
&  $1+\frac{r+1}{2}$\hfill \\ \hline
$D_r$, $r$ odd & $(0,\ldots,0,k)$  & $2\sum_{i=1}^{(r-1)/2}
H^{2i-1}+(2\!-\!r)H^{r-1}-rH^r$
&  $2$\hfill \\ \hline
$D_r$, $r$ odd & $(k,0,\ldots,0)$  & $2H^{r-1}+2H^r$
&  $1+\frac{r+1}{2}$\hfill $(*)$\\ \hline
$D_r$, $r$ odd & $(0,\ldots,0,k,0)$  & $2\sum_{i=1}^{(r-1)/2}
H^{2i-1}-rH^{r-1}+(2\!-\!r)H^r$
&  $2$\hfill $(*)$\\ \hline
$G_2$ & --     & 0  & $2$\hfill    \\ \hline
$F_4$ & --     & 0   & $3$\hfill    \\ \hline
$E_6$ & --     & 0   & $3$\hfill    \\ \hline
$E_6$ & $(0,0,0,0,k,0)$     & $H^1 - H^2 + H^4 - H^5$   & $3$\hfill $(*)$
\\ \hline
$E_7$ & --     & 0         & 4 \hfill   \\ \hline
$E_7$ & $(0,0,0,0,0,k,0)$    & $H^4 + H^6 + H^7$         & 2 \hfill   \\ \hline
$E_8$ & --     & 0        & 3 \hfill   \\ \hline
    \end{tabular}
\caption{Currents, Conjugacy classes and the number of O-planes.
 Column two gives
the Dynkin labels of the current (at level $k$) that corresponds to
the conjugacy class generator in column three according to
(\ref{eq:Qc}).
An asterisk indicates orientifold configurations related to a previous
one by an even current, in which case they are just centre
translations of each other.}
\label{tab:O-planes}
\end{center}
\end{table}

\begin{table}[ht]
\begin{center}
\small
    \begin{tabular}{|l||l|l|l|} \hline
$G$ & $n$       & $G_c$              & Dimension
\\ \hline \hline
$SU(N)$ & 0     & $S(U(p)\times U(q))$, $q$ even      &  $2pq$\hfill
\\ \hline
$SU(N)$ & 1     & $S(U(p)\times U(q))$, $q$ odd      &  $2pq$\hfill
\\ \hline
$SO(N)$, $N$ odd & 0     & $SO(p)\times SO(q)$, $q=0\mod 4$   &$pq$  \hfill
\\ \hline
$SO(N)$, $N$ odd & 1     & $SO(p)\times SO(q)$, $q=2\mod 4$     &$pq$  \hfill
\\ \hline
$SO(N)$, $N=2\mod 4$  & 0     & $SO(p)\times SO(q)$, $q=0\mod 4$   &$pq$  \hfill
\\ \hline
$SO(N)$, $N=2\mod 4$  & 1 (s) & $U(N/2)$  &$\frac{1}{4}N(N-2)$  \hfill
\\ \hline
$SO(N)$, $N=0\mod 4$  & 0     & $SO(p)\times SO(q)$, $q=0\mod 4$   &$pq$  \hfill
\\ \hline
$SO(N)$, $N=0\mod 4$  & 1 (s) & $U(N/2)$  &$\frac{1}{4}N(N-2)$  \hfill
\\ \hline
$SO(N)$, $N=0\mod 4$  & 1 (v) & $SO(p)\times SO(q)$, $q=2\mod 4$ & $pq$ \hfill
\\ \hline
$Sp(N)$ & 0     & $Sp(p)\times Sp(q)$   & $pq$\hfill
\\ \hline
$Sp(N)$ & 1     & $U(N/2)$     &$\frac{1}{4}N(N+2)$   \hfill
\\ \hline
$G_2$ &    0  & $SU(2) \times SU(2)$  & 8    \\ \hline
$F_4$ &   0   &SO(9)   & 16   \\ \hline
$F_4$ &   0   &$Sp(6) \times SU(2)$   & 28   \\ \hline
$E_6$ & 0     & $SO(10)\times U(1)$   & 32    \\ \hline
$E_6$ & 0     & $SU(6)\times SU(2)$   & 40    \\ \hline
$E_7$ & 0     & $SO(12)\times SU(2)$         & 64 \hfill   \\ \hline
$E_7$ & 1    & $E_6 \times U(1)$         & 54 \hfill   \\ \hline
$E_7$ & 1    & $SU(8)$         & 70 \hfill   \\ \hline
$E_8$ & 0     & $SO(16)$        & 128\hfill   \\ \hline
$E_8$ & 0     & $E_7 \times SU(2)$        & 112\hfill   \\ \hline
    \end{tabular}
\caption{Dimensions of O-planes. The first column indicates the
group manifold, the second the power $n$ of the current, and the third
the subgroup $G_c$ that commutes with the orientifold map. The dimension
of the O-plane is equal to the dimension of the tangent space of $G/G_c$
given in column 4. In all cases $p+q=N$. For $SO(N)$, $N$ even,
we indicate
in column 2 the conjugacy class under consideration.
}
\label{tab:O-planedims}
\end{center}
\end{table}

Next, consider the orientifold projection by $\Omega_n$ where $n$ is
even,
i.e. $n=2m$. Note that, when the order of the centre $N$ is odd, we can
always bring $n$ to this form. It is easy to see that~(\ref{eq:loc}) is
solved by
\beq
    g_{2m,u} = \c^m g_{0,u} \;\;\; ,
\eeq
i.e. the locations of the $n=2m$ planes are simply
translations of the canonical
($n=0$) ones by centre elements $\c^m$. It is not hard to see that
translation by an element of the centre is consistent with conjugation,
i.e. that
the conjugacy classes of $g_{0,u}$ map one-to-one to the conjugacy
classes of $g_{2m,u}$. Hence the $n=2m$
configuration of O-planes directly follows from the $n=0$ result.

For $N$ even, we have the additional possibility that $n=2m+1$
odd. Then
\beq
    g_{2m+1,u} = \c^m g_{1,u} \;\;\; ,
\eeq
solves~(\ref{eq:loc}) where $g_{1,u}$ is a solution for $n=1$. So the
locations of the $n=2m+1$ planes are translations of the $n=1$
configuration
by centre elements $\c^m$. The $n=1$ and $n=0$ configuration
appear to be related by translations
\begin{equation}
\label{eq:nis1}
g_{n=1,u}= \sqrt{\c}g_{n=0,u}~,
\end{equation}
where $\sqrt{\c}$ is one of the
elements of the maximal torus that squares to $\c$.\footnote{We may
  restrict our attention to elements of the maximal torus, since we
  are solving $g= \c^ng^{-1}$ only modulo conjugation. In fact
  eq.~(\ref{eq:nis1}) would not be a solution otherwise.} Of course
$\sqrt{\c}$ itself is not an element of the centre.
We must check anew how the $2^r$ solutions $g_{n=1,u}$ split into
conjugacy classes. The total number of O-planes is now given by the
number of Weyl orbits containing the $2^r$ solutions $g_{1,u}$. Again
we determined these numbers empirically and they are also given in
table \ref{tab:O-planes}.

Note that for
 $N$ even, a given configuration (both
$n=0,1$) is invariant under translations by the $\ZZ_2$ subgroup of the
centre. Indeed, from~(\ref{eq:loc}) it is easy to see that when $g$ is
 a
solution, so is $-g$.

\subsection{Dimensions of O-planes} \label{sec-dim}

To determine the dimension of the O-planes we need to know which
elements of the group commute with $g_{n,u}$, the element of the
maximal torus that defines the conjugacy class. This analysis can be
done in the tangent space at the point $g_{n,u}$ of the group manifold, in
other words we have to determine which elements of the Lie algebra
commute with $g_{n,u}$. For a generic element of the maximal torus,
this set (called the commutant $G_c$ henceforth)
consists of only the Cartan sub-algebra, and then the
dimension of the O-plane is $d-r$, where $d$ is the dimension of the
group and $r$ its rank. For a centre element the commutant
is of course the entire group, and therefore the O-plane
has dimension 0 (i.e. it is an O0-plane) if and only if $g_{n,u}$
is an element of the centre. In all other case we have to examine the
commutator of the root generators $E_{\alpha}$ with $g_{n,u}$.

This can be done as follows. Note that
\beq
E^{\alpha}[(t,H)]^m =
(t,H)\{E^{\alpha}-[(t,H),E^{\alpha}]\}[(t,H)]^{m-1}
                    =[(t,H)-(\alpha,t)]E^{\alpha}[(t,H)]^{m-1}~.
\eeq
Iterating this we get
\beq
E^{\alpha}[(t,H)]^m
                    =[(t,H)-(\alpha,t)]^m E^{\alpha}~,
\eeq
and hence
\beq
E^{\alpha}e^{2\pi i(t,H)}
                    =e^{2\pi i[(t,H)-(\alpha,t)]} E^{\alpha}~.
\eeq
Therefore $E^{\alpha}$ commutes with $e^{2\pi i(t,H)}$ if and only if
$(\alpha,t)$ is an integer.
Obviously linear combinations of root generators
are in the commutant if and only if
every term separately is in the commutant. Hence the commutant is in fact
a regular subalgebra.
The dimension of the plane is $d-r-M$, where
$M$ is the number of roots $\alpha$ with $(\alpha,t)$ integer. One may
also write this as $d-d_c$, where $d_c$ is the dimension of the
commutant $G_c$.
In other words, the dimension of the O-plane is equal
to the dimension of the coset space $G/G_c$.

We can say a little more about these coset spaces.
Since $[e^{2\pi i(t,H)}]^2$ is an element of the centre, it must commute
with all $E_{\alpha}$, and therefore $2(\alpha,t)$ must be an integer. Hence
the set of root generators splits into two sets, one with $(\alpha,t)$ even that
commutes with $e^{2\pi i(t,H)}$, and one with   $(\alpha,t)$ odd that anti-commutes
with it. This means that the Lie-algebra $\bg$ of $G$  has a direct sum  
decomposition
$\bg=\bg_c+\bar\mathfrak{k}$, where $\bg_c$ is the Lie algebra of the commutant and
$ \bar\mathfrak{k}$ is the `odd' part. Note that the commutator of two elements
of  $ \bar\mathfrak{k}$ commutes with $[e^{2\pi i(t,H)}]^2$,  and hence
is in $\bg_c$. This implies that the coset $G/G_c$ is a symmetric space.

The converse is also true: let $G_c \subset G$ be a regular embedding,
such that $G/G_c$ is a symmetric space. Then the Lie algebra of $G$ can
be decomposed as a direct sum $\bg=\bg_c+\bar\mathfrak{k}$, and there
exists an automorphism that fixes $\bg_c $ and sends $\bar\mathfrak{k}$
to minus itself. If the embedding is regular, $\bg_c $ contains the
entire Cartan subalgebra of $\bg$, and hence the Cartan subalgebra of
 $\bg$ is left unchanged by the automorphism. This implies that the
automorphism is inner, i.e. there exists an element $x \in G$ such that
$xhx^{-1}=h; xkx^{-1}=-k$ for $h\in \bg_c$ and $k\in \bar\mathfrak{k}$.
This element is unique up to multiplication by an element of the centre.
Furthermore $x^2$ commutes with both $h$ and $k$, and hence it is an
element of the centre.
Therefore $x$ is a solution to (\ref{eq:loc}).

Hence the classification of O-planes is now a matter of inspecting
the table of symmetric spaces \cite{SymSpac},
and selecting the ones where the
embedding is regular.
The
results are summarized in table
\ref{tab:O-planedims}.
In the table the trivial case, $G_c=G$, corresponding
to O0-planes is omitted. This is always a solution for $n=0$, and
corresponds to $g=1$. This table was actually
obtained by counting the number of odd
root generators in order to identify which conjugacy class belongs
to which symmetric space. The relation between conjugacy classes
and regular symmetric spaces is not one-to-one because of the possibility
of multiplying $x$ with a center element. This may or may not change
the conjugacy class to which $x$ belongs. For example, in the trivial
case $G=G_c$ (yielding O0-planes) valid choices for $x$ are 1 and $-1$
(if $-1$ is an element of $G$), but 1 and $-1$ are obviously not conjugate.
Up to this freedom of multiplying $x$ by elements of the centre,
the number of different choices of $G_c$ for given $G$
and $n$ corresponds to the number of O-planes in the last column of
\ref{tab:O-planes}. In most cases where there are more conjugacy
classes per symmetric space this degeneracy is due to multiplication
by $-1$, but
for $D_r$, $r$ even, there are
additional degeneracies.

This can be illustrated rather nicely for $G=SU(N)$, because in that case
one can solve (\ref{eq:gdef}) explicity in the group. For $n=0$ the
condition $g^2=1$ combined with $\det(g)=1$ allows diagonal matrices
with $p$ eigenvalues 1 and $q$ eigenvalues $-1$, with $N=p+q$ and $q$ even.
This breaks $SU(N)$ to $S(U(p) \times U(q))$. Therefore the
corresponding O-planes have dimension $2pq$.

\def\diag{{\rm diag}}
For $n=1$ we may take a square root of the center element
$\diag(\exp(2\pi i/N))$. If we take all $N$ entries of the square
root equal to $\exp(\pi i/N)$ we would get determinant $-1$. Hence we
have to flip an odd number of signs. The corresponding group element
$g$ breaks $SU(N)$ to $S(U(p) \times U(q))$, but this time with
$q$ odd. Note that if $N$ is odd either $p$ is odd or $q$ is odd, so that
there is no real difference between $n=0$ and $n=1$, as expected.

Note that O-planes have smaller dimensions than D-branes except if $G_c$ is
precisely the Cartan sub-algebra. This happens only for $SU(2)$, $n=1$.

\bigskip

To summarize, simple geometric semiclassical analysis of the WZW-action tells us
that the O-plane characteristics depend on the order of the centre.
When the order of the centre $N$ is odd, there is only one
configuration of O-planes modulo centre translations. This configuration has
one O$0$-plane at the origin; the remaining O-planes are localized on
conjugacy classes of various dimensions; their number is given in table
\ref{tab:O-planes}.
For even $N$, there are two inequivalent
configurations of
O-planes modulo centre translations. The `even' configuration has two
O$0$-planes at $g=\pm 1$, and a remainder --- see table
\ref{tab:O-planes} --- of planes
localized on conjugacy classes, whereas the `odd' configuration
consists completely of  O-planes localized on
 certain conjugacy classes. This
pattern generalizes to the non-cyclic centre of $D_r,~r$ even,
in a straightforward manner. When there are more cyclic factors in the
centre, the number of
inequivalent configurations is $2^s$, where $s$ is the number of even
cyclic factors.

\subsection{Planes in Orbifolds}

As for free CFT's, the WZW action (\ref{action}) may also be used to
describe orbifolds.
For every subgroup $H$ of the centre, we can construct the orbifold
$G/H$.
Let $H=\ZZ_M$, with $M$ a divisor of $N$, be generated by $\xi :=
\c^{N/M}$. Then
the orbifold consists of equivalence classes ($m=0,...,M-1$)
\beq \label{eq:orb}
    g \sim \xi^m g~.
\eeq
The $n=0$ O-planes in orbifolds are located at those points of $G/H$ for
which
\beq
    g = g^{-1}  \;\;\;\; .
\eeq
By~(\ref{eq:orb}), this relation only has to hold modulo translations by
$\xi^m$. In terms of the covering group $G$, every O-plane configuration
corresponding to $\xi^m$ appears in the orbifold. We will see in
section~\ref{sec-plorb} that the planes form $H$-translation invariant
combinations. Obviously this generalizes directly to $n\neq 0$ planes.

\subsection{An example: $SU(2)$}

Let us illustrate these results with
$SU(2)$.\footnote{This example was studied prior to our work
by C. Bachas et. al. \cite{Bachas}.} The group manifold is a three-sphere;
the conjugacy classes are two-spheres of constant latitude
and the centre elements are at the north and south poles. The standard
orientifold symmetry $g \rightarrow g^{-1}$ is a reflection through
the `axis of rotation' with fixed points at the poles. In terms of the
solutions~(\ref{eq:nis0}) we simply have $g_{u=0} = 1$ and $g_{u=1}=-1$.

The action of the centre is to identify antipodal points on the sphere. So the
combined action $g \rightarrow -g^{-1}$ leaves the equator
fixed. The solution~(\ref{eq:nis1})
indeed describes one O-plane at the
equator, since $\pm \sqrt{\c} = \pm $diag$(\i,-\i)$.

So the orientifold group $(1,\Omega_0)$ gives two O$0$-planes at the
north and south pole and the orientifold group $(1,\Omega_1)$ gives one
O$2$-plane at the equator.

When we identify $g \sim -g$, we obtain the group manifold of $SO(3)$. The
orientifold projection $g \rightarrow g^{-1}$ gives the following
configuration. The two $SU(2)$ O$0$-planes at the poles are identified by the  
orbifold symmetry, leaving one O$0$-plane at the pole for $SO(3)$. There is
also one O$2$-plane at the equator, since this submanifold is now fixed under  
the combined action of the orientifold and orbifold symmetry.

\medskip

Recalling the caveat regarding differing views of target space between
conventional string compactifications and WZW models, it is
interesting to briefly compare the $SU(2)$ results to the well known results
for $U(1)$; for, the $U(1)$ theory at the
self-dual radius is equal to the $SU(2)$
level one model. In terms of $U(1)$ the standard orientifold symmetry
$g \rar g^{-1}$ is just the reflection $x \rar -x$ modulo $2\pi R$, with
fixed points
$x=0$ and $x=\pi R$ corresponding to $g=\pm 1$.
The $n=1$ orientifold symmetry $g \rar -g^{-1}$
corresponds to the identification $x \simeq -x+ \pi R$ modulo $2\pi R$, with fixed
points at $x=\pm \pi R/2$. Within $SU(2)$ these solutions are conjugate, as
are all solutions to the conditions $g^2=\gamma$, $\gamma\in U(1)$: the
center is the entire group.
If we only consider the locations there is therefore just one solution
in $U(1)$, up to equivalence. However, if we also consider the tensions
there are two orientifold configurations
for any radius $R$, one with O0-planes
with opposite tensions and one with O0-planes
with the same tension \cite{SagP}. In the case of $U(1)$ at the self-dual radius
or $SU(2)$ level 1, the underlying CFT interpretation of these two
configurations is the same, and the difference corresponds in both cases
to a simple current modification of the crosscap, as discussed in the
next section, eqn. (\ref{eq:cross}). In section 5 we will see that the two
O0-planes in $SU(2)$
actually have opposite tension. Hence this configuration corresponds
to the analogous one in the $U(1)$ picture. The other configuration
is, in $SU(2)$ language, an O2-plane with non-zero tension. This corresponds
in $U(1)$ language to two O0-planes with the same tension. In both cases
the latter configuration is the one that one obtains from the
simple current modified crosscap (\ref{eq:cross}) with $n=1$.

\setcounter{equation}{0}

\section{Stringy analysis} \label{sec-stringy}

The above geometric analysis is only valid in the large volume limit,
i.e. at infinite level. At finite $k$, we expect our results to
acquire stringy corrections. Conformal field theory, being exact in
the string tension, should provide these corrections.

We therefore consider the chiral algebra $\cal{A}$ of a WZW model.  It
contains both the Virasoro algebra and an affine Lie algebra
$\mathfrak{g}$
spanned by the modes of a chiral current $W$ given by ($n,m\in \ZZ$
and $a,b=1,...,d$)
\beq \label{eq:affine}
    [W_n^a,W_m^b] = f^{ab}_c W^c_{n+m} + m \delta^{ab} \delta_{n,-m} K
    \;\;\;\; .
\eeq
Here $K$ is the central element whose eigenvalue is the level $k$ and
$f^{ab}_c$ the structure constants of the
horizontal Lie algebra $\bg$. At finite level, WZW
models are rational (i.e. the number of unitary highest weight
representations is finite). The unitary highest
weight representations of $\cal{A}$ are labeled by $\L=(\bL,k)$ where $\bL$
is a highest weight representation of the horizontal subalgebra
$\bg$ such that $(\bL,\Psi)\leq k$ with $\Psi$ the
highest root of $\bg$. The set of highest weights $\bL$ which obey
this constraint, form the dominant affine Weyl chamber $\cW^+_k$.

For closed oriented theories
the chiral and anti-chiral algebras have to be combined in such a way
that the torus partition function
\beq \label{eq:T}
    \cZ= \sum_{\L,M} \chi_{\L} Z^{\L,M} \tilde{\chi}_{M}
\eeq
is modular invariant. Here $\chi_{\L}$ are the affine
$\mathfrak{g}-$characters of a representation $\L$. For modular invariance,
it is necessary and sufficient to
require the matrix $Z$ to commute with the modular matrices $S$ and
$T$; explicit expressions for these matrices are known in the case of
WZW models and are given in the appendix.
The invariant $Z$ corresponds to the geometry of the space on
which the strings live. When we take for $Z$ the charge conjugation
invariant, this space is the compact, simply-connected Lie group $G$
whose Lie algebra is the compact real form of $\bg$. With the
use of simple currents we can construct a large class of inequivalent
invariants.
Simple current invariants correspond geometrically to orbifolds of $G$ by
(subgroups of) the centre~\cite{strWZW,invariants}. For
instance, for $A_1$ level $k$, the modular invariant choices for $Z$
come in an ADE
classification. The A-series correspond to the charge conjugation
invariant (which is just the diagonal invariant in this case); it
describes an $SU(2)$ manifold. The D-series are simple current
invariants that describe $SO(3)$. Finally, the E-series, that only
exist for three values of $k$, do not correspond to a simple current.

Recall that simple currents~\cite{simple} are primary fields whose fusion rules
are such that fusion with {\em any} other primary field yields
exactly one other field and not several.
The fusion rules of simple currents among
themselves form an Abelian group. This group is
isomorphic to the centre of
$G$.\footnote{Except for $E_8$ level 2 \cite{Fuchs}.}
We will denote the field that generates the
simple current group $H$ by $J$, the
elements by $J^n$, $n=0,...,N-1$ and the identity by $0$, i.e. the
vacuum representation. An important quantity is the monodromy charge,
defined by ($h_\L$ is the conformal weight of $\L$ and $J\L$ is the unique
fusion product of $J$ and $\L$.)
\beq \label{eq:charge}
    Q_J(\L) := h_J + h_{\L} - h_{J\L} \mod 1 \;\;\; .
\eeq
The charges of the other currents in $H$ are related to that of the
generator by $Q_{J^n}(\L) = n Q_J(\L)$ mod $1$. The monodromy charge is
related to the conjugacy class $\cC$ of the $\bg$-representation $\L$ as
\beq \label{eq:Qc}
    Q_J(\L) = \frac{\cC(\L)}{N} \mod 1 \;\;\; .
\eeq

\medskip
To construct an unoriented theory, we
project the spectrum described by~(\ref{eq:T}) to states that
are invariant under the interchange of left- and rightmovers by adding
the Klein
bottle partition function~\cite{sagnotti}
\beq \label{eq:K}
    \cK = \sum_{\L \in \cW_k^+} K^{\L} \chi_{\L}~.
\eeq
To ensure that the total closed unoriented
spectrum, encoded in $(\cT+\cK)/2$, produces nonnegative state
multiplicities, $K_{\L}$ must be an integer
that has the same parity and is bounded
by $Z_{\L,\L}$. The perturbative worldsheet expansion in this theory
contains unoriented surfaces, i.e. surfaces with crosscaps.
The worldsheet may also have boundaries. Their
properties are encoded in the annulus and M\"{o}bius strip partition
function:
\begin{equation}
  \label{eq:6}
  \cA_{ab} = \sum_{\L \in \cW_k^+} A^\L_{ab} \chi_\L~,~~~~
\cM_a = \sum_{\L \in \cW_k^+} M_a^{\L} \chi_\L~.
\end{equation}
The Chan-Paton labels $a,b$ are used to distinguish the possible
boundaries of the open strings. Open-closed string duality means that
we may equivalently describe the three partition functions $\cK,~\cA$
and $\cM$ as the tree-level ``exchange'' of closed strings,
propagating between boundary-states $|B_a\rangle$ and crosscap states
$|\Gamma\rangle$:
\beq \label{eq:C}
    |B_a \rangle = \sum_{\L \in \cW_k^+} B_{a,\L}
    |\L\rangle\rangle_{B}~,
~~~|\Gamma \rangle = \sum_{\L \in \cW_k^+}
    \Gamma_{\L} |\L \rangle\rangle_{\Gamma}~.
\eeq
We have expanded each onto a basis of
Ishibashi boundary-
and crosscap
states respectively in order to make the conformal properties of these
states manifest; $B_{a,L}$ and $\Gamma_{\L}$
are known as the {\em
  boundary} and \emph{crosscap coefficients}. These are not arbitrary
but must obey various integrality
constraints.
For all simple current modifications of the charge
conjugation invariants solutions to these
constraints are
known~\cite{FOE}. In case of the charge conjugation invariant, we have
$N$
solutions; one for
every simple current $J^n$:
\begin{equation} \label{eq:cross}
  B_{a,\L}^{(n)} = \frac{S_{a,\L}}{\sqrt{S_{J^n,\L}}}  \;\;\;\;, \;\;\;\;
\Gamma^{(n)}_{\L} = {P_{J^n,\L} \over
      \sqrt{S_{J^n,\L}}}~,~~~~~n=0,\ldots, N-1~,
\eeq
where $S_{a,b}$ is the modular $S$-matrix and $P_{a,b}$ the `pseudo'-modular
matrix equaling
$P:=\sqrt{T}ST^2S\sqrt{T}$. This
matrix will form an important part of our story, and we will
discuss it in more detail below. Historically, the boundary state with the
$n=0$ coefficient is the famous Cardy state. The crosscap coefficient with
$n=0$ first appeared
in~\cite{planar}. In~\cite{klein}, it is shown that the above set of
boundary and crosscap coefficients
gives rise to a consistent unoriented spectrum for every simple current
$J^n$. Not all currents give rise to a different spectrum however.
In~\cite{thesis}, it is shown that two `Klein bottle currents' $J^n$ and
$J^{n'}$ produce the same spectrum if $n+n'$ even modulo $N$.

We claim that, just as the boundary states describe stringy D-branes,
these crosscap states describe the
orientifolds found in section~\ref{sec-semi} in the stringy regime. As
a check, we would like to recover the semiclassical properties and
positions in the
large volume limit. In order to extract the position of the defects
described by the
crosscap state, we follow the method of~\cite{brWZW, DiVec}, by calculating
the coupling of a `graviton' to the crosscap in
`momentum space'.
Extrapolating from the $U(1)$ free CFT and its chiral
extension
 $U(1)_k$, a `graviton' is
created by the vertex
\begin{equation}
W^a_{-1}\tilde{W}^b_{-1}+W^b_{-1}\tilde{W}^a_{-1}|\mu,\tilde{\mu}\rangle_M
\end{equation}
The weights $\mu,\tilde{\mu}$ are the
left- and right-`momentum', and required to be weights of $M$ and $M^c$
respectively (since we are discussing the charge conjugation modular
invariant). Due to the multi-dimensionality of the
ground-state of the Verma-module (by acting with the horizontal
subalgebra), the range of $\mu$ is not just the set of affine highest weight
states $\cW_k^+ =\{\L \in L_w | (\L,\Psi) < k\}$, but includes all
weights $\l$ found by acting with the horizontal subalgebra on $\cW^+_k$.
To find the location of the orientifold we
compute the one-point function of a graviton,
up to a normalization of the crosscap states which will be discussed
in section 5.
\begin{equation}
G^{ab}_M(\mu,\tilde{\mu})={}_M\langle \mu,\tilde{\mu} | W^a_1
\tilde{W}^b_1+W^b_{1}\tilde{W}^a_{1}  |\Gamma \rangle~.
\end{equation}

We will need the explicit form and reflection properties of the
Ishibashi crosscap state
\begin{equation}
 \label{eq:twish}
W^a_m | \L \rangle\rangle_{\Gamma} = (-1)^m\tilde{W}^a_{-m}| \L
 \rangle\rangle_{\Gamma}~,~~~~
|\L \rangle\rangle_{\Gamma} =\sum_{{\rm
 levels }~m} |\L,m\rangle \otimes U
 (-1)^m |\tilde{\L}=\L,m\rangle~,
\end{equation}
where $U$ is an anti-unitary operator that maps weights $\m \in M$ to their
duals $\m^c\in M^c$. Here
$|\L,m\rangle$ is a condensed notation for the complete set of
weights of the horizontal algebra $\bg$ that appear at level $m$ in
the Verma module based on the highest weight $\L$. With the help of
(\ref{eq:twish})
one finds that
\begin{eqnarray}
\label{ert}
G^{ab}_M(\mu,\tilde{\mu})&=& -{}_M\langle \mu,\tilde{\mu}| W^a_1W^b_{-1}
+W^b_{1}{W}^a_{-1}|\Gamma \rangle \nonumber
\\
&=& - 2 k \delta^{ab} \sum_{\L}
\Gamma_\L \, {}_M\langle \mu,\tilde{\mu} | \L \rangle\rangle_{\Gamma} =
-2k\delta^{ab} \delta_{\tilde{\m},\m^c} \Gamma_M~.
\end{eqnarray}
It is obvious from the computation that $-2 k\delta^{ab}$ just
signifies the tensorial nature of the overlap, and that one may use
any arbitrary closed string state to derive the momentum-space distribution
(e.g. as in \cite{Malda-etal}). We will therefore drop this factor
from now on, and call the remainder $G_M(\mu,\tilde{\mu})$.

To find the spatial-distribution we need to ``Fourier'' transform. The
analogue of Fourier modes for group-manifolds are the matrix elements
$\langle\l|R_{\L}(g)|\l'\rangle$, where $R_{\L}(g)$ stands for the group
element $g$ in
the representation $\L$ and $\l,~\l'$ are weights
$\l \in \L$ and $\l'\in \L^c$. The Fourier transformation to find the
distribution amounts to
computing
\begin{equation}
  \tG (g) = \sum_{\L} \sum_{\l,\l'} G_\L(\l,\l') \langle\l|R_{\L}(g)|\l'\rangle
= \sum_{\L
  \in \cW_k^+} \cX_{\L}(g) \Gamma_{\L}~.
\label{pppp}
\end{equation}
The last step follows from the fact that highest weights $\L \notin
\cW^+_k$ do not appear in the theory.

The above is simply repeating what has been done for the boundary
states in \cite{brWZW}. There
on the rhs of (\ref{pppp}) the boundary coefficient, and thus the modular
$S$-matrix appears instead of the crosscap coefficient which depends on
the modular $P$-matrix. Mathematically it was known that
the modular $S$-matrix is expressible
in terms of a character of the horizontal Lie group. This meant that the boundary
state analogue of (\ref{pppp}) was easily evaluated by using
completeness of the group
characters, and the geometric
analysis could proceed straightaway. Our aim in the next section will
be to derive an expression for the $P$-matrix in
terms of group characters of $G$ as well.

\setcounter{equation}{0}
\section{The $P$-matrix, characters and Gaussian sums}

\subsection{Properties of the $P$-matrix}

Before we turn to this calculation of the $P$-matrix in WZW models,
we briefly
discuss some of its properties that hold for arbitrary CFT's. As we indicated,
the $P$-matrix plays an important role
in the construction of open unoriented string
theories~\cite{sagnotti}. The $P$-matrix
encodes the {\em channel transformation} for the M\"obius strip, i.e.
it relates the open and closed string channels of this diagram~\cite{sagnotti}.
The action $P$
may be represented in terms of the modular $T$- and $S$-matrix as
$P=\sqrt{T}ST^2S\sqrt{T}$,
with $T$ diagonal and $\sqrt{T}$ defined as ${\rm exp}[i\pi (h_a-c/24)]$.

When the CFT has a $\ZZ_N$ simple current group generated by $J$, the matrix
elements of $S$ are related as follows ($n=0,...,N-1$)~\cite{simple, intril}:
\beq
    S_{J^ni,j} = e^{2\pi\i n Q_{J}(j)} S_{i,j}~,   \label{eq:SJ}
\eeq
where $Q$ is the monodromy charge~(\ref{eq:charge}).
A similar relation can
be derived for the matrix elements of the $P$-matrix~\cite{FOE}
\beq \label{eq:PJ}
    P_{J^{2m}i,j} = \phi (2m,i) e^{2\pi\i mQ_{J}(j)} P_{i,j}~,
\eeq
where
\beq \label{eq:sigma}
\phi (2m,i) := \exp\left[\pi\i\left(h_i - h_{J^{2m}i}
-2Q_{J^m}(J^mi)\right)\right]~.
\eeq
In the application in the main text, we have $i=0$ or $i=J$ and $J$
integer spin, in which case the phases
$\s(2m):=\phi(2m,0)$,~$\s(2m+1):=\phi(2m,J)$ are
just signs.

Note that there is no simple relation for matrix elements that are
related by odd currents. This paper basically explains why: the
corresponding O-plane configurations are not related by a centre
translation.

\subsection{The $P$-matrix in WZW models and $G$-characters}

The objective of this section, to which we now return,
is to find an expression for the $P$-matrix of
WZW models in terms of group characters, since this enables us to find the
location of the O-planes by using orthogonality of these characters.

For WZW models we know the explicit expressions for both the modular
$S$- and the modular $T$-matrix. We have given them in the
appendix. The
$P$-matrix is therefore explicitly known. Some
properties of Lie algebra
characters and the different forms in which they may be
expressed are also given in the appendix. Inspection shows that to find
$P$
in terms of characters we must perform the matrix multiplication inherent in the
definition of $P$.
After substituting the formulas for the $S$ and $T$ matrix
we therefore have to compute
\begin{eqnarray}
    && \hspace{-0.3in}P_{\L,N} =  \cN_{S}^2 \exp\left[-\pi\i
 \left(\frac{k d}{4h}  \right) \right] \exp{\left[
 \frac{\pi\i}{2h}
 \left((\L,\L + 2 \rho)+(N,N + 2 \rho)
 \right ) \right] }  \\ \nonumber
&& \times
 \sum_{\stackrel{\scriptstyle{R \in \cW_k^+}}{w,v \in
 W}} {\rm sign}
(wv) \exp\left[ \frac{-2\pi\i}{h} \left[
    (w(\L+\rho),R+\rho) - (R,R + 2 \rho) +
    (v(N+\rho),R+\rho)\right] \right]~.
\end{eqnarray}
Here $\cW_k^+$ is the set of affine dominant integral weights, i.e. the set of  
weights $R$ of $G$ with nonnegative Dynkin labels such that
$(R,\Psi) \leq k$, with $\Psi$ the
highest root. The vector $\rho$ is the Weyl vector,
$h:=k+g^{\vee}$ the
height, $g^{\vee}$ the dual Coxeter number, $W$ the Weyl group
of $G$ and sign$(w)$ equals $(-1)^{{\rm\#~generators~of~} w \in W}$.
$\cN_S$ is the
normalization of the $S$-matrix given in
eq. (\ref{eq:hjs}).

We can simplify this expression by shifting the sum over $R$ to a sum
over $M=R+\rho$.
When $(R,\Psi) \leq k$ then $(R+\rho,\Psi) \leq k + (\rho,\Psi)= h - 1$.
Furthermore, as the Dynkin labels $R_i\geq 0$, it follows from the
definition of the Weyl vector that
$R_i+\rho_i=R_i+1\geq 1$. So this shifted
weight $M$ lives in the interior of the strictly dominant
affine Weyl chamber $\cW_{h}^{++}$.  However, since the $S$-matrix vanishes
at the boundaries of this chamber,
we can extend the sum back to the dominant Weyl
chamber $\cW_h^+$.
\begin{eqnarray}
&& \hspace*{-0.6in}
P_{\L,N}=\cN_{\cS}^2 \exp{\left[ -\pi\i
 \left(\frac{k d}{4h}  \right ) \right ] } \exp{\left[
 \frac{\pi\i}{2h}
 \left( (\L,\L + 2 \rho)+ (N,N + 2 \rho) -4(\rho,\rho)\right)
 \right ] }
\\ \nonumber
&&\times  \sum_{\stackrel{\scriptstyle{M \in \cW_h^{+}}}{w,v \in W}}
{\rm sign} (wv) \exp\left[ \frac{-2\pi\i}{h} \left[
    (w(\L+\rho),M) - (M,M) + (v(N+\rho),M)\right] \right]~.
\end{eqnarray}
Next, for every $v$ we reorder the sum over $w$ to $vw$. Then, since
Weyl reflections preserve the inner product, we can combine the sums
over $M$ and $v$ to a sum over weights $M$ that live in all Weyl
reflections of the dominant Weyl chamber, i.e. the set $\cW_h$. It may
appear that the fixed points of the Weyl group at the boundaries are now
undercounted. However, as we have seen before, they don't contribute to the
$P$-matrix, so we can add or subtract them at will.

There is, however, a more natural range for $M$.
Note that the summand is invariant under a
translation $M \rightarrow M +h\a$ with $\a$ an arbitrary simple
coroot.
Recall that for every
element of the weight lattice $L_w$ that does not lie on the boundary of
some Weyl chamber, there is a unique element of the affine Weyl group (at
`level' $h$) that maps it to the dominant Weyl chamber $\cW_h^+$. The affine
Weyl group is the (semi-direct)
product of the horizontal Weyl group $W$ and the group of
translations by $hL^\vee$, with $L^\vee$ the coroot lattice. So every weight
in $\cW_h = W(\cW_h^+)$ that does not lie at the boundary of some chamber,
represents one of the elements in $\tilde{L}_w/hL^\vee$, where $\tilde{L}_w$
is the weight lattice where all boundaries are removed. So instead of summing  
over all elements in $\cW_h$, we may sum over cosets $M \in L_w/hL^\vee$,
where we included the boundaries again since they do not contribute to $P$
anyway.
So
\begin{eqnarray}
P_{\L,N}&=&
\cN_{\cS}^2 \exp{\left[ -\pi\i
 \left(\frac{k d}{4h}  \right ) \right ] } \exp{\left[
 \frac{\pi\i}{2h}
 \left( (\L,\L + 2 \rho)+ (N,N + 2 \rho) -4(\rho,\rho)\right)
 \right ] }
\\ \nonumber &\times&
\sum_{\stackrel{\scriptstyle{M \in L_w/hL^\vee}}{w\in W}} {\rm
  sign} (w) \exp\left[ \frac{-2\pi\i}{h} \left[
    (w(\L+\rho),M) - (M,M) + (N+\rho,M)\right] \right]~.
\end{eqnarray}
We can now complete squares in the second line:
\begin{eqnarray}
 P_{\L,N} &=&
\cN_{\cS}^2 \exp{\left[ -\frac{\pi\i}{4h}
 \left(k d + 12(\rho,\rho)  \right ) \right ] }
\\ \nonumber &\times&
\sum_{\stackrel{\scriptstyle{M \in L_w/hL^\vee}}{w\in W}}
\exp\left[ \frac{-\pi\i}{h} \left(
    w(\L+\rho),N+\rho\right) \right] \exp\left[
  \frac{2\pi\i}{h} \left(M - \frac{w(\L+\rho) +
      N+\rho}{2}\right)^2 \right]~.
\end{eqnarray}
Upon using the strange formula for the length of $\rho$
(recall that we are using the
normalization $(\Psi,\Psi)=2$)
\begin{equation}
  \label{eq:14}
 (\rho,\rho) = \frac{1}{12} dg^{\vee},
\end{equation}
we can write this as
\beq
\label{pln}
P_{\L,N}= \cN_{\cS}^2 \exp{\left[ -
 \frac{\pi \i d}{4}  \right ] } \sum_{w\in W} {\rm sign}(w)
\exp\left[ \frac{-\pi\i}{h} \left(
    w(\L+\rho),N+\rho\right) \right]  \cG_h (C^{-1},w(\L+\rho) + N+\rho)
\eeq
where we introduced the lattice Gaussian sum on $L_w$ with
metric $C^{-1}$.
\begin{equation}
\cG_h (C^{-1},X) :=  \sum_{M \in L_w/hL^\vee} \exp\left[
    \frac{2\pi\i}{h} \left|M - \frac{X}{2}\right|^2
  \right]~.
\end{equation}
In the next section, see equation~(\ref{eq:GS}), we prove that this sum equals
\beq
\label{ffrst}
\cG_h(C^{-1},X) = \left|{L_w \over L^\vee}\right|^{1/2} \left(h\i\over 2
\right)^{r/2} \sum_{u\in L^\vee/2L^\vee} (-1)^{\frac{h}{2}(u,u) +
(u,X)}~,
\eeq
where
$u$ is the set of all coroots modulo even coroots.
In other
words, $u$ is the set of vectors $u=\sum_i u_i\tilde{\a}^i$ on the
coroot
lattice
$L^\vee$ whose entries $u_i$
are either $0$ or $1$. Hence $(u,X)= \sum_i u_iX^i$ where $X^i$ are
the Dynkin labels, and $(u,u)=\sum_{j=1}^r u_i C^{ij}u_j$ where
$C$ is the inner product matrix of the coroots $C^{ij} =
(\tilde{\a}^i,\tilde{\a}^j)$.

Taking everything together, we find
\beq \label{eq:P}
P_{\L,N} = \cN_\cP \sum_{u \in L^{\vee}/2L^{\vee}}
(-1)^{\frac{h}{2}(u,u) + (u,N +
  \rho)} \sum_{w\in W} {\rm sign}(w) \exp\left[
  \frac{-\pi\i}{h} \left( w(\L+\rho),N+\rho + hu
       \right ) \right]
\eeq
with
\beq
\cN_\cP := 2^{-r/2} \cN_\cS = \i^{(d-r)/2} |L_w/L^\vee|^{-1/2}
\frac{1}{\sqrt{(2h)^r}}~.
\eeq

Two results immediately follow from eq. (\ref{eq:P}). Using the Weyl
character formula (\ref{weylch})
$P$ can be seen to equal the sum of
characters $\cX_\L$ of the horizontal subgroup,
\beq
\label{pchc}
P_{\L,N} = \sum_u \cN_u^{(N)} \cX_\L(g_u^N)
\eeq
evaluated at the group element
\beq
g_u^N = \exp\left\{\frac{-\pi \i}{h}
(N+\rho+hu, H)\right\}~,
\eeq
where $\cN_u^{(N)}$ equals
\beq
 \cN_u^{(N)} := \cN_P (-1)^{\frac{h}{2}(u,u)+(u,N+\rho)} \sum_{w \in W} {\rm
sign}(w) \exp \left\{ \frac{-\pi \i}{h} (w(\rho), N+\rho+hu)\right\}~.
\eeq
And, due to the formal equivalence between characters $\cX_\L$ and the
modular $S$ matrix, one derives a linear relation between $S$ and $P$~,
\beq
P_{\L,N} = 2^{-r/2}\sum_u (-1)^{\frac{h}{2}(u,u)+(u,N+\rho)}
S_{\L,\frac{N-\rho+hu}{2}}~.
\eeq
This relation exists only in the formal sense, as the $S$-matrix for
non-integer weights is only defined by analytic continuation
through its equivalence with
characters. Nonetheless the very existence of these relations is
remarkable. Both, however, hinge on the proof of eq. (\ref{ffrst})
with which we
now proceed.

\subsection{Gaussian sums for Lattices}

The Gaussian sum $g_k(a,0)$ is the discrete analog of the Gaussian integral:
\begin{equation}
  \label{eq:28}
  g_k(a,0)=\sum_{n=0}^{k-1} e^{\frac{2\pi i a n^2}{k}}~.
\end{equation}
Gaussian sums play a role in analytic number theory (for a review,
see~\cite{Gsums}). Their main
application is to count the number of solutions to the congruence
$x^2 \equiv a (\mbox{mod}~p)$; the number
$a$ is called a quadratic residue mod $p$.

This sum has been evaluated to
\begin{eqnarray}
  \label{eq:29}
  g_{mk}(a,0)&=& mg_k(a,0)~,~~~~\mbox{gcd}(ma,mk)= m \e \ZZ~, \non
  g_k(a,0) &=& \left( \frac{a}{k}\right) g(1,k)~,~~~~\mbox{gcd}(a,k)=1 ~,\non
  g_k(1,0) &=& \frac{1}{2}(1+i)(1+i^{-k})\sqrt{k}~,
\end{eqnarray}
where $\left(\frac{a}{k}\right)$ is the Legendre symbol. Decomposing
$k$ into primes $k=\prod p_i$, it equals the
product of Jacobi symbols,
\begin{eqnarray}
  \label{eq:30}
  \left(\frac{a}{k}\right)&=& \prod_{i;k=\prod
  p_i}\left(\frac{a}{p_i}\right)~,
\non \left(\frac{a}{p}\right) &=&\left\{ \matrix{ 1 &\mbox{if $x^2=a
  (\mbox{mod}~p)$ has a solution for $x \e \ZZ$} \cr
      -1&\mbox{if not}}\right.~.
\end{eqnarray}

For purposes of comparison, we also give the Gaussian sum over half integers
\begin{eqnarray}
  \label{eq:31}
  g_k(a,\frac{1}{2})&=&\sum_{n=\frac{1}{2}}^{k-\frac{1}{2}}
  e^{\frac{2\pi i a n^2}{k}}
  = \frac{1}{2} g(a,4k)-g(a,k)~.
\end{eqnarray}
For $a=1$ this is simply
\begin{equation}
  \label{eq:32}
  g_k(1,\hlf)=\frac{1}{2}(1+i)(1-i^{-k})\sqrt{k}~.
\end{equation}
For $SU(2)$ the weight lattice is one-dimensional and the sum in
 (\ref{pln}) equals a regular Gaussian sum. For higher rank
groups we will need a lattice generalization of this sum.

Consider an even lattice $L^\vee$ with volume $|L^\vee|$ and dual lattice
$L_w$. In the application we have in mind, $L^\vee$ is the coroot lattice of
a Lie algebra $\bar{\mathfrak{g}}$ and $L_w$ the weight lattice, but the
following discussion applies to arbitrary even lattices. The restriction to
even lattices is important, since only then is $L^\vee$ (and every integer
multiple $hL^\vee$) a sublattice of $L_w$. The quadratic Gaussian sum on
$L_w/hL^\vee$ is defined as
\begin{eqnarray}
    \cG_h (C^{-1},X) &:=&  \sum_{\m \in L_w/hL^\vee} \exp\left\{
    \frac{2\pi\i}{h} \left| \m - \frac{X}{2}\right|^2
  \right\} \\
&=& \sum_{m^i} \exp\left[
    \frac{2\pi\i}{h} (m^i-\frac{X^i}{2})C^{-1}_{ij}(m^j-\frac{X^j}{2})
  \right]~,
\end{eqnarray}
where $X$ is an integral weight and $|y|^2$ is the length squared of
$y$, computed with the lattice metric $C^{-1}$ of $L_w$.
The precise
summation
range of $m_i$ follows from the first form of the expression.
In order to evaluate this sum, it is convenient to define ($\e > 0$)
\beq
\label{reg}
    \tau := h - \i\e \;\;\; \rightarrow \;\;\; {1\over \t} = \frac{1}{h}
+\frac{\i\e}{h^2} +\cO(\e^2)~.
\eeq
Consider then the sum over the full weight lattice
\beq \label{eq:GaussInf}
    \T(\t) := \sum_{\l \in L_w} \exp\left\{\frac{2\pi\i}{\t} \left|\m -
\frac{X}{2}\right|^2
  \right\}~,
\eeq
which due to the small regulating parameter introduced in (\ref{reg})
is now finite and well defined (it is a theta function).
We can write $\T$ in two different ways. In the first, we split the sum over
$\l$ in a $\e$-dependent sum over $h\b$, $\b \in L^\vee$ and a
$\e$-independent sum over\footnote{We have to
be careful here, since the summand is not independent of the representative
of the coset $\m \in L_w/hL^\vee$. We must therefore choose a particular set
of representatives and this is implicit in the notation. In the final
answer, equation~(\ref{eq:GS}), this ambiguity disappears.} $\m \in
L_w/hL^{\vee}$:
\beq
\T(\t) = \sum_{\m \in L_w/hL^\vee} \exp{\[{2\pi\i\over
h} \left|\m - {X \over 2} \right|^2\]}
\sum_{\b\in L^\vee}  \exp{\[ -2\pi\e
\left|\b+\frac{\mu-X/2}{h}\right|^2+\cO(\e^2)\]}~.
\eeq
The last sum can be approximated by a Gaussian integral
that to leading order in $\e$ does not depend on $\m - X/2$, and we
recognize the Gaussian sum in the prefactor:
\beq \label{eq:split}
    \T(\t) = \cG_h(C^{-1},X) \[ {1 \over \sqrt{\e}} \int dx e^{-2\pi x^2} +
\cO(1) \]~.
\eeq
On the other hand, we can first
use Poisson's resummation formula (which is
allowed since $\e >0$) to write the sum~(\ref{eq:GaussInf})  as
\beq
   \T(\t) = |L_w|^{-1}  \({\i\t \over 2}\)^{r/2}  \sum_{\a \in L^\vee}
\exp{\[\frac{-\pi\i \t}{2}|\a|^2\]} \exp{\[2\pi\i({X \over 2},\a)\]}~,
\eeq
and then
split~\footnote{Again, at this stage we must choose a representative
for $u$ in order for this expression to be well-defined.}
 the sum over $\a$ in a $\e$-dependent sum over $2\b, \b \in L^\vee$
and a $\e$-independent sum over $u \in
L^\vee/2L^\vee$
\beq
\T(\t) =  |L_w|^{-1}  \({\i\t\over 2}\)^{r/2}  \sum_{u \in L^\vee/2L^\vee}
(-1)^{(u,X) +
\frac{h}{2}(u,u)}\sum_{\b \in L^\vee} \exp{\[-2\pi\e|\b - \frac{u}{2}|^2\]}~.
\eeq
As before, we can approximate the sum over $\b$ by an integral that does not
depend on $u/2$ to leading order in $\e$:
\beq \label{eq:pois}
    \T(\t) = |L_w|^{-1}  \({h\i\over 2}\)^{r/2}  \sum_{u \in
L^\vee/2L^\vee}  (-1)^{(u,X) +
\frac{h}{2}(u,u)} \[ {1 \over \sqrt{\e}} \int dx e^{-2\pi X^2} + \cO(1) \]~.
\eeq
We can now equate the two results~(\ref{eq:split}) and~(\ref{eq:pois}).
Comparing the $\cO(1/\sqrt{\e})$ terms, we find for the quadratic Gauss sum
\beq
\label{eq:GS}
    \cG_h(C^{-1},X) = \left|{L_w \over L^\vee}\right|^{1/2}  \({\i
      h\over 2}\)^{r/2}
\sum_u (-1)^{(u,X) +
\frac{h}{2}(u,u)}~,
\eeq
where we used $|L_w| = |L^\vee|^{-1}$ in obtaining the final equation.
For rank one $SU(2)$ this indeed reduces to the known
results~(\ref{eq:30}) and (\ref{eq:32}).

To appreciate how remarkable this result is note that we started
with a summation of gaussians involving the integrable weights
at a certain level. Yet the result depends on the level in a
very simple way, namely only through the volume of the affine Weyl
chamber.
Although not shown here explicitly, the dependence
on $X$ is also very simple. A priori it is clear that the result
only depends on the classes $L_w/2L_w$, and is constant on the Weyl
orbits of these classes (the number of distinct classes is in fact
precisely given by the entries in the last column of table \ref{tab:O-planes}).
If $h$ is even, the summation over $u$ yields $2^r$ for $X$=0, and zero if
$X$ belongs to a non-trivial class. For odd $h$ the result is slightly
more complicated. The summation can take three values on the classes,
either 0 or $\pm 2^l$, with $l \leq r$.
There are many more interesting features we could mention, and
undoubtedly more still to be discovered, but let us return now to
the main subject of this paper.

\setcounter{equation}{0}
\section{The semiclassical limit}

We are now in a position to show that there is a one-to-one correspondence
between
the O-planes constructed in section~\ref{sec-semi} and the  crosscap states
of section~\ref{sec-stringy}. We will do this by demonstrating that the
distribution~(\ref{pppp}),
that gives the location of the defects described by the
crosscap states in the large volume limit, is peaked at the solutions
$g_{n,u}$ of equation~\R{loc}. Let us first consider the charge conjugation
invariant.
There is a normalization issue we have suppressed up till
now. In the more extensively studied case of D-branes, the
graviton-crosscap coupling is not directly given by
the boundary coefficients, but by a reflection coefficient $R_{\Lambda,a}$
that
differs from it by a factor~\cite{branes, CarLew}. By ``boundary
coefficient" we concretely
mean here the coefficients $B$ that appear in the expression of the
annulus as $A^i_{ab}=\sum_{\Lambda} S^i_{\Lambda}
B_{{\Lambda}a}B_{{\Lambda}b}$, and that are
determined by integrality. These are the coefficients presented in
(\ref{eq:cross}). The additional factors come from the definition of the
annulus amplitude and the normalization of the identity near a boundary.
In the Cardy case, the reflection coefficients are $S_{{\Lambda}a}/S_{0a}$,
and the boundary coefficients $S_{{\Lambda}a}/\sqrt{S_{{\Lambda}0}}$.
We should expect similar factors to appear between the crosscap
states $\Gamma_{\Lambda}$ (\ref{eq:cross}) and
the O-plane reflection coefficients. The latter we
will denote as $U_{\Lambda}$. An additional complication is the fact that
we have to know these factors
for non-zero values of $J$. Our proposal is to use the following
definitions
\begin{equation} \label{eq:cross4}
  R_{a,\L}^{(n)} = \frac{S_{a,\L}}{{S_{0a}}}  \;\;\;\;, \;\;\;\;
U^{(n)}_{\L} = {P_{J^n,\L} \over
      S_{00}}~,~~~~~n=0,\ldots, N-1~,
\eeq
It may appear counter-intuitive that for $n\not=0$ the phases from
the denominator in (\ref{eq:cross}) were dropped. The rationale behind this
is that the D-brane reflection coefficients should not change if we
modify the O-plane reflection coefficients, by going from $P_{0\L}$ to
$P_{J^n,\L}$. On the other hand, at the one-loop level, the M\"{o}bius
strip amplitude does change, and this in its turn enforces (through
open sector integrality) a change in the annulus amplitude. The latter
change correponds always to a different
choice of the boundary
conjugation matrix $A^0_{ab}$. This can be taken into account
by extra phases in the boundary coefficients. In other words,
the phases in the denominators in (\ref{eq:cross}) belong naturally in the
boundary and crosscap coefficients that enter in the definition of the
one-loop partition functions, but not in the reflection coefficients.
Obviously this is a conjecture, and not a proof.
The fact that this definition will turn out to make sense geometrically
gives additional support for this conjecture, and we hope that it
may help resolve a long-standing debate on the proper description and
the origin of these phases. In fact, we will see that the geometric
picture gives a nice way of understanding boundary conjugation.

The precise distribution that we therefore
wish to evaluate is
\beq \label{eq:tG}
    \tG^{(n)} (g) = \sum_{\L \in \cW_k^{+}} U_{\L}^{(n)} \cX_{\L} (g)
\eeq
rather than eq. (\ref{pppp}).
In the infinite volume limit, $k \rightarrow \infty$, the sum is over all
highest weight representations of $\bar{\mathfrak{g}}$. We can then use
completeness of the group characters, provided we have an expression for
$P_{J^n,\L}$ as a sum of $G$-characters. This is the very fact we
accomplished in (\ref{pchc})~,
\beq \label{eq:pchar}
P_{J^n,\L} = \sum_u \cN^{(n)}_u \cX_{\Lambda} (g^{(n)}_u)~,
\eeq
where
\beq \label{eq:Nu}
 \cN^{(n)}_u := \cN_\cP (-1)^{\frac{h}{2}(u,u) + (u,J^n + \rho)} \sum_{w\in
W} {\rm sign}(w) \exp\left\{-\frac{\pi\i}{h}
(w(\rho),J^n+\rho + hu) \right\}~,
\eeq
and the character in the $P$-matrix is evaluated at the group element
\beq \label{eq:g}
    g^{(n)}_u := \exp\left\{ -\frac{\pi\i}{h}
      (J^n + \rho,  H)  \right\}
    \exp\left\{-\pi\i (u,H)  \right\}~.
\eeq
We will start with $n=0$. In the infinite volume limit, the argument of
the characters in~\R{pchar} becomes (we ignore $\rho/h$)
\beq
  g_{u}^{(0)} \stackrel{h \rightarrow
\infty}{\longrightarrow} e^{-\pi\i(u,H)} = g_u~,
\eeq
where $g_u$ is the solution to~\R{loc} for $n=0$.
To work out the normalizations we make
use of the denominator identity,
\beq
\sum_{w \in W} {\rm sign}(w) e^{(w(\rho),\lambda)} =
\prod_{{\rm
positive~roots}~\a} 2\sinh\left[\frac{(\a,\lambda)}{2}\right] \ ,
  \eeq
applied to the ratio
\beq
 {\sum_{w\in
W} {\rm sign}(w) \exp\left\{-\frac{\pi\i}{h}
(w(\rho),\rho+hu ) \right\}
\over
 {\sum_{w\in
W} {\rm sign}(w) \exp\left\{-\frac{2\pi\i}{h}
(w(\rho),\rho ) \right\} }}~.
\eeq
The result is
\beq
 \prod_{{\rm
positive~roots}~\a} {\sin\left[ {\pi\over 2h} (\a,\rho+hu)\right]
\over
\sin\left[ {\pi\over h} (\a,\rho)\right]}~.
\eeq
If $(\alpha,u)$ is even, the $u$ term may be dropped, and in the limit
$h\to\infty$ the corresponding factor in the product contributes a factor
${1\over2}$. This is obviously the correct answer if $u=0$:
\beq
    \frac{\cN_0^{(0)}}{S_{0,0}} \rightarrow
\cQ_0^{(0)} :=2^{-r/2}2^{-(d-r)/2} = 2^{-d/2}\ ,
 \eeq
If on the other hand $(\alpha,u)$ is odd (since $u$ is a co-root
it is integer) the numerator contributes
$\sin({1\over 2} \pi(\alpha,u))$, which
remains finite in the limit $h\rightarrow\infty$.
We can write the limit of the ratio as follows
\beq
{\sin\left[ {\pi\over 2h} (\a,\rho+hu)\right]
\over
\sin\left[ {\pi\over h} (\a,\rho)\right]} \rightarrow
e^{\frac{1}{2}\pi\i (\alpha,u)} \times
\cases{ \frac{1}{2}& for $(\alpha,u)$ even \cr
{h\over \pi\i(\alpha,\rho)}& for $(\alpha,u)$ odd\cr }
\eeq
The result is then
\beq
    \frac{\cN_u^{(0)}}{S_{0,0}} \rightarrow
\cQ_u^{(0)} := 2^{D/2}(-1)^{\frac{h}{2}(u,u)}\left[{  h \over \pi\i
}\right]^{D/2}
\prod_{\a, {\rm odd}} \left[{1\over
(\a,\r)}\right]\cQ_0^{(0)},
 \eeq
where the product is over all positive roots with $(\a,u)$ odd, and
where $D$ is the dimension of the O-plane. This is equal to
$D=d-r-M$, where $M$ is the number of roots with $(\a,u)$ even. The
number of ``odd" positive roots is equal to $\frac{1}{2}(d-r)-\frac{1}{2}M=
\frac{1}{2}D$. Since we started from a real expression, the result is
of course real. This implies that $D$ must be a multiple of four, as indeed
it is according to table \ref{tab:O-planedims} (with $n=0$!).
Note that
the result diverges for large $h$ as
$ h^{D/2}$.
Obviously this is a volume factor. At finite $h$, previous results
on D-branes suggests that the branes are fuzzy. Hence even an O0 plane,
a point, has then a finite volume. But as $h$ tends to infinity, the
ratio of the volume of a higher dimensional object and that of a point
goes to infinity. This phenomenon has not been observed for WZW
D-branes for the simple reason that they all occupy regular conjugacy
classes of the same dimension $d-r$
\cite{brWZW}. Although we believe we correctly understand the relative
normalization, the overall normalization is another matter, as is the
relative normalization between D-branes and O-planes.
Overall normalizations of amplitudes are tricky in general, and even more
so in our case where one does not consider a genuine string theory, so that
`graviton scattering' is only called that way by analogy.
We will not try
to address this issue here and turn instead to the main goal, the
localization of the planes, for which the precise normalization is
of little relevance.

The distribution~\R{tG}
can now easily be computed using completeness of the group characters:
\bea
    \tilde{G}^{(0)}(g) & = & \sum_u \cQ_u^{(0)}  \sum_{\L}
    \cX_{\L} (g_u) \cX_{\L}(g) \non
    & = & \sum_u \cQ_u^{(0)}\d(g_{u}-g) \;\;\; .
\eea
This shows that the
crosscap state $|\Gamma^{(0)}\rangle$ corresponds to a localized charge
distribution in curved geometries and that it provides a conformal field
theory description
of the $n=0$ orientifold planes of section~\ref{sec-semi}. Note that,
since the Weyl group preserves inner products, the contributions to the
distribution from characters $\cX_\L$ evaluated at
$g_u$ and $g_{w(u)}$ have the same
coefficient $\cQ_u^{(0)}$. This is what one should
expect, since this factor can be interpreted as a charge
(in fact a tension) and $g_u$ and
$g_{w(u)}$ describe the same plane, as we discussed in
section~\ref{sec-semi}.
The delta function is actually a $\delta$-function
on conjugacy classes, not on individual group elements. The sum on $u$
usually contains several members of the same conjugacy class, thus
producing a multiplicity factor which one may absorb in the
definition of $\cQ_u^{(0)}$.
Note that the distribution we get
is peaked at {\it all} locations of O-planes corresponding to $n=0$, in
contrast to the situation for D-branes, which live on a single conjugacy
class. This reflect the wel-known fact that D-branes come with Chan-Paton
multiplicities which can be switched on or off
at will, whereas there is no such
freedom for O-planes: one can either choose to have one or not to have one, but
one cannot distribute them freely over the available locations.

In a similar way, the crosscap states $|\Gamma^{(n\neq
  0)}\rangle$ describe
O-planes that correspond to a non-trivial element of the centre. First take
$n$ even, $n=2m$. Now (note that $\s(2m)$ is a sign, see the discussion
after equation~(\ref{eq:sigma}))
\bea
    P_{J^{2m},\L} & = & \s(2m) e^{2\pi\i Q_{J^m}(\L)} P_{0,\L}
    \non
    & = & \s(2m) \sum_u  \cN^{(0)}_u e^{2\pi\i \frac{m\cC(\L)}{N} }
    \cX_{\L} (g^{(0)}_{u}) \non
    & = & \s(2m) \sum_u  \cN^{(0)}_u \cX_{\L} (\c^m g^{(0)}_{u})~,
\label{steps}
\eea
where we used~(\ref{eq:Qc}) and~(\ref{eq:charC}). In the large volume limit,
$g^{(0)}_u \rightarrow g_u$ and the distribution becomes
\beq
    \tilde{G}^{(2m)}(g) =  \s(2m)  \sum_u \cQ_u^{(0)} \d(\c^mg_{u}-g)~.
\eeq
So the corresponding O-plane configurations are translations of the canonical
ones by $\c^m$, up to an overall sign. When we
compare this with the discussion in
section~\ref{sec-semi}, we conclude that the crosscap state
$|\Gamma^{(2m)}\rangle$ describes the O-plane configuration with
centre element $\c^{2m}$ in the defining orientifold group.

Now the computation for $n=1$, $N$ even, remains.
Following the same steps in eq. (\ref{steps}),
we can show that for $n=2m+1$ the crosscap reflection
coefficient is
\beq
    P_{J^{2m+1},\L}  =  \s(2m+1) \sum_u \cN^{(1)}_u \cX_{\L} (\c^m
    g^{(1)}_u)~.
\eeq
Note that in the
large $h$ limit, ${J^n/h}$ is finite. Hence
in the large volume limit, the group element~\R{g} approaches, for $n=1$,
\beq
    g^{(1)}_u \rightarrow \exp\left\{-\pi\i\frac{1}{N} \sum_{i=1}^r c_i H^i
\right\}
 \exp\left\{-\pi\i(u,H)  \right\} = \sqrt{\c} g_{u}~,
\eeq
where $\sqrt{\c}$ is a group element that squares to the generator of the
centre $\c$. In obtaining this result, we have written $(J,H)$ in Dynkin
basis as $\sum_{ji} J_j C^{-1}_{ji} H^i$. The Dynkin labels of a simple
current are always of the form $J_j=k\delta_{qj}$ for some $q$.
Now we observe
the following property of the
quadratic form matrix $C^{-1}$: for all simple Lie algebra's the
column vector $C^{-1}_{iq}$ equals
\beq \label{eq:Cl}
    C^{-1}_{iq} = \frac{c_i}{N} \mod 1~,
\eeq
as can be shown by simple inspection of the tables in for
instance~\cite{books}. From this relation between the quadratic form matrix
and the conjugacy
classes, it follows that $\sqrt{\c}$ indeed squares to $\c$ and that
$g^{(1)}_u$ approaches the solutions~\R{nis1} in the large volume
limit. To compute the tension in the large volume limit we need to consider
\bea
    \cQ_u^{(1)}  &:=  &\lim_{h\rightarrow\infty}
\frac{\cN_u^{(1)}}{S_{0,0}} = \lim_{h\rightarrow\infty}
\frac{\cN_u^{(1)}}{S_{0,J}} \\ \nonumber
&=& \lim_{h\rightarrow\infty}2^{-r/2} (-1)^{\frac{h}{2}(u,u) + (u,J+\rho)}
 \prod_{
\a > 0} {\sin\left[ {\pi\over 2h} (\a,\rho+J+hu)\right]
\over
\sin\left[ {\pi\over h} (\a,\rho+J)\right]}
\eea
The $J$ in the denominator is introduced just for convenience. At level
$k$ the simple currents can be written as $k J_1=(h-g^{\vee})J_1$, where  
$J_1$ is a
fundamental weight.
The ratio of the sines can then be written as follows
\beq
{\sin\left[ {\pi\over 2h} (\a,\rho+J+hu)\right]
\over
\sin\left[ {\pi\over h} (\a,\rho+J)\right]}
={
{ \exp({\frac{1}{2}\pi\i(\a,J_1+u)}) \left[
\exp({ {\pi\i\over 2h} y}) - \exp({\pi\i(\a,J_1+u)})
\exp({-{\pi\i\over 2h} y}) \right] }
\over
{ \exp({\pi\i(\a,J_1)}) \left[
\exp({ {\pi\i\over h} y}) -
 \exp({-{\pi\i\over h} y}) \right] }}
\eeq
where $y=(\a,\rho-g^{\vee}J_1)$. Using (\ref{eq:Cl}) the inner product
$(\a,J_1)$ equals
\beq
 (\a,J_1) = \sum_{i,j} \alpha_i C_{ij}^{-1} (J_1)_j =
 \sum_{i} \alpha_i c_i/N \mod 1
\eeq
Shifting the values of $c_i$ by $N$ does not affect the corresponding
centre element. Therefore it is possible to define $c_i$ in such a way
that the last relation holds exactly, not just modulo integers. Then
we get, using (\ref{eq:sol}), $(\a,J_1+u)=2(\alpha,t)$.  For the ratio
we find then
\beq
 e^{\frac{1}{2}\pi\i (\alpha,J_1+u)} e^{-\pi \i (\alpha,J_1)}\times
\cases{ \frac{1}{2}& for 2$(\alpha,t)$ even \cr
{h\over \pi\i(\alpha,y)}& for 2$(\alpha,t)$ odd\cr }
\eeq
Putting everything together and using $\rho=\frac{1}{2}\sum_{\a>0} \a$ we
find
\beq
    \cQ_u^{(1)}=2^{D/2}2^{-d/2}(-1)^{\frac{h}{2}(u,u) + (u,J)}e^{-i\pi(\rho,J_1)}
\left[{  h \over \pi\i
}\right]^{D/2}
\prod_{\a, {\rm odd}} \left[{1\over
(\a,\r-g^{\vee}J_1)}\right] \ ,
\eeq
where $\a, {\rm odd}$ means ``all positive roots with $2(\alpha,t)$ odd".
As in the case $n=0$ the tension diverges as $h^{D/2}$. In this
case $D$ is not always a multiple of 4, but the factor $e^{-\pi\i(\rho,J_1)}$
is purely imaginary precisely if $D=2\mod4$, so that the final result is
real.

The
distribution becomes
\beq
    \tilde{G}(g)^{(2m+1)} =  \s(2m+1)  \sum_u \cQ_u^{(1)}
    \d(\c^m\sqrt{\c} g_{u}-g) ~.
\eeq
This shows that $|\Gamma^{(2m+1)}\rangle$ describes the O-planes with
a nontrivial centre element $\c^{2m+1}$ in the defining orientifold
group. We have thus succeeded in showing that
the geometric interpretation of simple current crosscap states
 for the charge
conjugation invariant of a WZW model are the centre O-plane configurations
on a compact, simply-connected Lie group manifold. The exact CFT
computation has also given us the (relative) charges of each.

\subsection{Planes in Orbifolds} \label{sec-plorb}

We will now consider a WZW model with a simple current modular
invariant. To
be more precise,
we will only consider invariants that are extensions
of the
chiral algebra~\cite{simple}. Such an invariant can be built with the use
of an
integer
spin (conformal weight) simple current group $H$, that is a subgroup
of the
centre. We take $H=\ZZ_M$, where $M$ a divisor of $N$ and denote the
generator by $L := J^{N/M}$. Only fields that have monodromy charge
zero with
respect to all currents in $H$ contribute to the
torus partition function. The geometric interpretation of this
invariant is that of strings living on an orbifold $G/H$. Only
representations of $G$ that are not-faithful with respect to the
conjugacy
classes corresponding to $H$ appear as representations of this
coset. Given
the relation between the simple current charges and conjugacy classes,
equation~\R{Qc}, it follows that all representations of $G/H$ appear
in the
torus partition function.

For simple current
modifications of the charge conjugation invariant,
the
crosscap coefficients are~\cite{FOE}
\beq \label{eq:cross2}
    \Gamma_{\L} = \frac{1}{\sqrt{|H|}} \sum_{n=0}^{M-1} \h(n)
    {P_{L^n,\L}
\over \sqrt{S_{0,\L}}}  \;\;\; .
\eeq
The signs $\h(n)$ are constrained by the requirement that $\C_{\L}$
vanishes
for charged fields:
\beq \label{eq:sign}
    \h(n) = \h(n+2m) \exp[\pi\i(h_{L^{n}} - h_{L^{n+2m}})] \;\;\;\; ,
\eeq
for all $m=0,...,M$.

We now want to give a geometrical
  interpretation of the crosscap states for this theory, based on the
  crosscap coefficients in
equation~(\ref{eq:cross2}) and the solutions of
  the sign-rule~(\ref{eq:sign}). First note that the distribution
corresponding to this crosscap state
  is still given by~(\ref{eq:tG}), simply because the crosscap coefficient
vanishes for those
  $\L$ that are not
representations of the orbifold when the $\eta$ satisfy the
sign-rule. Again, we must make a distinction between the crosscap
  coefficient
and the graviton-crosscap coupling. Following our previous argument, we presume
the last is given by:
\beq
    U_{\L} = \frac{1}{\sqrt{|H|}}
\sum_{n=0}^{M-1} \h(n) \frac{P_{L^n,\L}}{S_{00}}   \;\;\; .
\eeq
Let us first consider $M$ odd. We can rewrite the sum over $P$-matrix
elements as follows:
\bea
    \sum_{n=0}^{M-1} \h(n) P_{L^n,\L} & = & \sum_{m=0}^{M-1} \h(2m)
    P_{L^{2m},\L} \non
     & = & \eta(0) \sum_{m=0}^{M-1} e^{2\pi\i Q_{L^m} (\L) } P_{0,\L} \non
    & = & \eta(0) \sum_{m=0}^{M-1} \sum_u \cN^{(0)}_u \cX_{\L} (\c^m
    g^{(0)}_{u})~.
\eea
In the first step we used that all currents in $\ZZ_{\rm odd}$ groups
are even. In the second step, we combined the signrule~(\ref{eq:sign})
and the identity~(\ref{eq:PJ}).
In the last step, we used our character formula~(\ref{eq:P})
for the $P$-matrix and used
equation~(\ref{eq:charC}) from the appendix.
In the limit $h\rightarrow \infty$, the group
element $g^{(0)}_u \rightarrow g_u$, so the distribution $\tG$ becomes
\beq
    \tG (g) = \frac{\eta(0) }{\sqrt{|H|}} \sum_{m=0}^{M-1} \sum_u
\cQ^{(0)}_u  \d(\c^m
    g_{u}-g)~.
\eeq
Hence this crosscap state describes an $H$-translation invariant
combination that survives the orbifold projection. The overall sign
$\eta(0)$ can be interpreted as the overall sign of the charges
(tensions) of the O-planes.

For $M$ even, we have to make a distinction between even and odd currents:
\bea
    \sum_{n=0}^{M-1} \h(n) P_{L^n,\L} & = & \sum_{m=0}^{M/2-1} \h(2m)
    P_{L^{2m},\L} + \sum_{m=0}^{M/2-1} \h(2m+1) P_{L^{2m+1},\L}
    \\
    & = & \eta(0) \sum_{m=0}^{M/2-1} \sum_u \cN^{(0)}_u \cX_{\L} (\c^m
    g^{(0)}_{u}) + \eta(1) \sum_{m=0}^{M/2-1} \sum_u \cN^{(1)}_u \cX_{\L}
    (\c^m g^{(1)}_{u})) \nonumber~.
\eea
In the large volume limit, this crosscap state describes a configuration of
O-planes given by
\beq
    \tG (g) =\frac{\eta(0) }{\sqrt{|H|}}  \sum_{m=0}^{M/2} \sum_u
\cQ^{(0)}_u  \d(\c^m
    g_{u}-g) +  \frac{\eta(1) }{\sqrt{|H|}}  \sum_{m=0}^{M/2} \sum_u
\cQ^{(1)}_u  \d(\c^m\sqrt{\c}
    g_{u}-g)~.
\eeq
This describes two $H$-invariant combinations, if we take into account
that every $m$ in the sum describes a $\ZZ_2$ translation-invariant
configuration. The relative sign is actually the relative sign of the
tensions of the two $H$-invariant configurations, and is therefore relevant.

Similar remarks apply for general simple current invariants. The
crosscap state is always of the form~(\ref{eq:cross}). The number of
free signs equals the number of even cyclic factors $p$ in the simple
current group, so we have $2^p$ different crosscaps. In geometric
terms, this is the number of relative tensions between the $1+p$
translation invariant combinations of O-plane configurations.

\setcounter{equation}{0}
\section{Branes and Planes}

The geometry of D-branes in open oriented WZW models has been
discussed in~\cite{alSc,brWZW, stanciu,Malda-etal}. What we would like
to stress here, is
that this geometry differs in the \emph{un}oriented open
string. As reviewed in section~\ref{sec-stringy}, boundaries in CFT are
described by a boundary state $|B_a\rangle$. The boundary and crosscap states  
must be such that the open unoriented string partition function
$(\cA+\cM)/2$ produces nonnegative integers. Due to the M\"obius projection,
the
set of boundary labels $a$ is reduced; some boundaries are identified and
some boundaries are projected. More precisely, define the boundary
conjugate $a^*$ of $a$ by
\beq
    A^0_{ab} = \left\{ \begin{array}{cc}
                1 & {\rm when} \;\; b=a^* \\
                0 & {\rm when} \;\; b \neq a^*
                \end{array} \right.~,
\eeq
then complex boundaries $a\neq a^*$ are identified and real boundaries
are projected. In terms of Chan-Paton gauge groups, on a complex pair
of boundaries lives a unitary group. Real boundaries carry a
symplectic or orthogonal gauge group, depending on whether the
M\"obius coeffcient $M^0_a$ is plus or minus one.

In~\cite{FOE}, the boundary and crosscap coefficients (and therefore
the annulus and M\"obius strip coefficients) for arbitrary simple
current invariants are presented (see~\cite{thesis, walcher} for
proofs). We will discuss the charge conjugation invariant first. The
boundary labels $a$ are in one-to-one correspondence with
the primary fields (so we label them by $A$). Recalling
the boundary coefficients
corresponding to the crosscap
coefficients~(\ref{eq:cross})
\beq
    B^{(n)}_{\L,A} = \frac{S_{\L,A}}{\sqrt{S_{\L,J^n}}}~,
\eeq
we can calculate the boundary conjugation matrix
\beq \label{eq:conj}
    A^* = J^nA^c    \;\;\; .
\eeq
The geometric interpretation of these branes can be derived in a
similar way as was done in~\cite{brWZW}: calculate the brane-graviton
coupling and Fourier
transform this to a distribution on group space.
Again the coupling of a graviton
to a brane is given by the boundary {\em reflection}
coefficient~\cite{brWZW} $R_{\L,A}$, which differs from the boundary
coefficient by a normalization.
The reflection coefficient is exactly equal to a character of the
Lie group $G$
\beq
    R_{\L,A} = \frac{S_{\L,A}}{S_{0,A}} = \cX_{\L} (g_{A})
\eeq
evaluated at
\beq
    g_{A} = \exp\left\{\frac{-2\pi\i}{h} \sum_{i=1}^r
      ((A+\rho),H) \right\} \;\;\; .
\eeq
For a boundary $A$ the distribution that gives the localization of the
defect  in the limit of infinite level is given by
\bea
    \tilde{G}_{A} (g) & = & \sum_\L R_{\L,A} \cX_{\L} (g) \non
    & = & \sum_{\L}
    \cX_{\L} (g_{A}) \cX_{\L} (g) \non
    &  = & \d(g_{A} - g) \;\;\; .
\eea
where $g_A$ is the group element in the $h\rightarrow \infty$ limit:
\beq
    g_{A} \rightarrow \exp\left\{2\pi\i \sum_{i=1}^r (\a,
      H) \right\} \;\;\; ,
\eeq
where we keep the relative weight $\a^i = A^i/h \in$ R fixed in
taking the limit. To obtain
the location of the conjugate boundary, note that
\beq
    S_{\L,J^nA^c} = e^{2\pi\i Q_{J^n}(\L)} S_{\L,A}^*=
    e^{2\pi\i \frac{n}{N} \cC(\L)} S_{0A} \cX_{\L} (g^{-1}_{A})
    = S_{0A} \cX_{\L} (\c^n g^{-1}_{A})
\eeq
where we used two well-known properties of the $S$-matrix,
$S_{ij}=S_{ij^c}^*$ and equations~(\ref{eq:SJ}) in the first step, the
correspondence between monodromy charges and conjugacy
classes~(\ref{eq:Qc}) in the second and the property~(\ref{eq:charC})
in the last. The location of the conjugate boundary is therefore
\beq
    \tilde{G}_{A^*} (g) = \d(\c^n g^{-1}_{A} - g) \;\;\; .
\eeq
So the boundary conjugation~(\ref{eq:conj}) has a nice
geometric interpretation: given a boundary $A$ at $g_{A}$, its
charge conjugate is located at $R_{n}(g_A) = \c^ng_A^{-1}$. Real boundaries
therefore correspond to branes that are left fixed (not necessarily
pointwise) by the orientifold group, whereas complex boundaries form
an $R_n$ invariant pair.

Take again $SU(2)$. In the oriented theory, the boundaries are located
at circles of constant latitude. When we perform the standard
orientifold projection, i.e. reflection through the axis of rotation,
all these circles are left fixed. The branes either have an $SO$ or
$Sp$ gauge group, depending on the Frobenius-Schur indicator of the
boundary label $A$. When $A$ is a vector representation of $SU(2)$,
the projection is orthogonal and branes labelled by a spinor
representation are symplectic\footnote{Here, and in the following, we
  made the choice $-1$ for the overall sign of the M\"obius
  strip.}. When we gauge the orientifold group with a non-trivial
element of the centre, a brane is identified with its image obtained
by reflection through the equator. These branes carry unitary gauge
theories on their worldvolumes. Only for even level we have
a real boundary with an orthogonal gauge group, labelled by $A=k/2$,
that lies on top of the orientifold plane.

\subsection{Branes in orbifolds}

Boundary states for a (symmetric) simple current invariant are
presented in~\cite{FOE} (see~\cite{thesis} for a proof of integrality of the
annulus coefficients). We will again consider an extension
invariant by a group $H = \ZZ_M$ and generator $L=J^{N/M}$. We will assume that
the simple currents do not have fixed points (for a definition,
see~\cite{simple}). Then the boundary states are labelled by $H$-orbits
$[A]
:= (A,LA,L^2A,...,L^{M-1}A)$ and the Ishibashi states only exist for
chargeless fields $\L$. The reflection coefficients are given by
\beq
    R_{\L,[A]} = \sqrt{|H|} \frac{S_{\L,A}}{S_{0,A}} \;\;\; .
\eeq
The distribution to be calculated is
\beq
\tilde{G}_{[A]}  =  \sqrt{|H|}
    \sum_{\L, \cC_H(\L)=0} \cX_{\L} (g_{A}) \cX_{\L} (g)~.
\eeq
We restricted the sum over $\L$ to non-faithful $G$ representations with
respect to the subgroup $H$, since the boundary coefficient is not
automatically zero for chargeless fields. We can extend the sum to all $\L$
by inserting a projector
$$
\d^M_{\cC(\L),0} = \frac{1}{|H|} \sum_{m=0}^{M-1} e^{2\pi\i m
  \cC(\L)/M}~,
$$
where the superscript $M$ on the delta-function means modulo $M$. Then
\bea
    \tG_{[A]} & = & \frac{1}{\sqrt{|H|}}  \sum_{m=0}^{M-1}
    \sum_{\L} \cX_{\L} (\c^m g_{A}) \cX_{\L} (g) \nonumber \\
    & = & \frac{1}{\sqrt{|H|}}  \sum_{m=0}^{M-1} \d(\c^m g_{A} - g)~.
\eea
We used again the equivalence between characters of the Lie group and
the $S$-matrix, the relation between conjugacy classes and monodromy
charges and, in the last step, we inserted a projector on the trivial
class and used~(\ref{eq:charC}). From this result we immediately see
that the boundary states in the orbifold theory describe
$H$-invariant combinations.

When the simple currents that define the modular invariant partition
function do have fixed points, some boundary and Ishibashi labels may have
multiplicities. In the presence of O-planes, i.e. the crosscap given
by~(\ref{eq:cross2}), the boundary coefficients have an extra phase that depends  
on the multiplicity label and the relative O-plane charge (the signs $\eta$).  
(In~\cite{FOE} this phase is  $\a_J$.) Although it would be
interesting to give a geometrical interpretation of the multiplicities and
this phase, we will leave this for future research.

\section{Conclusion}

We have seen that crosscap states in WZW models and orbifolds thereof
correspond to localized orientifolds, as they
do in free CFT. The centre of the horizontal subalgebra plays an important
classifying role in the possible orientifold projections; this is the
geometrical counterpart of the role of simple currents in the exact CFT
description. To compute the various numbers, locations, dimensions and
charges of the possible geometric O-planes, we needed to prove
mathematically important lattice generalization of Gaussian sums. The
consequence of this is a highly remarkable relation between the
pseudo-modular $P$-matrix and characters of the horizontal subalgebra.

This
in turn implies a linear relation between the $P$- and the $S$-matrix.
It is intriguing that linear relations between the two matrices is
precisely what is needed in order to cancel tadpoles in open string
theories. It would be very interesting to know if such a relation
also exists in other CFT's. Although on the one hand there is no
obvious generalization of the analytic continuation of the labels of $S$,
on the other hand most CFT's are WZW-based. It is therefore conceivable
that such a relation might be derivable at least for coset CFT's. Whether
relations of this kind actually do help with tadpole cancellation is
an interesting question which we hope to address in the future.

\medskip

{\bf Acknowledgements:}
A.N. Schellekens wishes to thank C. Bachas for a discussion of his work on
the $SU(2)$ case. We thank B. Pioline for pointing out \cite{jeffrey}
in which lattice gaussian sums are considered. 
Although the method used to derive the
sum is essentially the same as ours, the gaussian sum considered in
\cite{jeffrey} is different; there the range is
over $L_w/hL_w$ instead of $L_w /h L^{\vee}$.

The work of K. Schalm was supported in part by
DOE grant DE-FG02-92ER40699.

\appendix

\setcounter{equation}{0}

\newcommand{\alinea}[1]{\subsubsection*{\normalsize \it #1}}
\section{Characters}
\label{characters:lie}

In this appendix, we list some formulas that are used in the main text. For
an extensive review on Lie algebras and groups, we refer to~\cite{books}.

A character maps the information of a highest weight
representation $R_\L$ of the Lie algebra
$\bar{\mathfrak{g}}$ to the complex numbers:
\beq \label{eq:character}
    \bar{\chi}_\L (h_t) = {\rm Tr} \exp\{2\pi\i R_\L(h_t) \}~,
\eeq
where $h_t=\sum t_i H^i \equiv (t,H)$ is an element of the Cartan subalgebra,
whose generators we denote as $H^i$.
The trace is over all weight vectors $\l$ in the representation
 with highest weight $\L$ and $R_{\L}(h_t)$ is the $\L$ representation of
$h_t$. As weights are eigenstates of the Cartan
subalgebra, $R_{\L}(H^i) \ket{\l}=
\l^i\ket{\l}$, we have
\beq
      \bar{\chi}_\L (h_t) = \sum_\l {\rm mult}_\L(\l) \exp\{2\pi\i(\l,t)\}~,
\eeq
where ${\rm mult}_\L(\l)$ denotes the
multiplicity of the weight $\l$ in the representation
$R_{\L}$. A second, more powerful, expression for the character
is given by the Weyl character
formula,
\beq
\label{weylch}
    \bar{\chi}_\L (h_t) = \frac{\sum_{w\e W} {\rm sign} (w)
      \exp{[2\pi i(w(\L+\rho),t)]}}{\sum_{w\e W} {\rm sign} (w)
      \exp{[2\pi i(w(\rho),t)]}}~.
\eeq
The sum is over the Weyl group $W$ weighted by the sign of each
      element defined as $\sign{w} =(-1)^{{\rm \#~of~generators~of}
      ~W}$. The vector $\rho$ is the Weyl vector and equals half the sum of
positive roots, or, equivalently, the sum
      of
      the
      fundamental weights $\rho=\sum_{i=1}^r \Lambda_i$.
These expressions are basis-independent. It is often convenient
to choose the Dynkin basis in weight space and to choose a basis in
the Lie-algebra (Chevalley basis) so that the eigenvalues of $H^i$ are
the Dynkin labels. The inner product in weight space is then
$(\lambda,\mu)=\sum_{i,j}\lambda_i C^{-1}_{ij} \mu_j$, where $C^{-1}$ is
the inverse of the symmetrized Cartan matrix.

\alinea{Group characters}

Eq.~(\ref{eq:character}) may also be read as a character of
 the Lie Group $G$,
 \begin{equation}
   \label{eq:9}
  \cX_\L(g) = {\rm Tr}_\L (g) ~,
 \end{equation}
evaluated at the group element $g_t = \exp\{2\pi \i\sum_i t_i H^i\}$. In the main  
text we use two properties. One is
that characters shift by a phase under
transformations $g \rightarrow \c g$ where $\c$ is the generator of the
centre of $G$:
\beq \label{eq:charC}
    \cX_\L (\c^n g) =  \exp\{2\pi\i \frac{n}{N} \cC (\L) \} \chi_\L
    (g) \;\;\; .
\eeq

What is really important to us, however, is that
characters serve as the analogue of Fourier modes for group
manifolds. They are orthogonal and complete
\beq \label{eq:orth}
\int \mu_G(g) \cX_\L^* (g) \cX_M (g) =\d_{\L,M}~,~~~
    \sum_{\L} \cX_\L(g) \cX_{\L} (h) = \d(g-h)
\eeq
The sum is over dominant weights only and the delta-function
$\d(g-h)$ is defined with respect to the Haar measure $\int
d\mu_G(a) \cF(a) \d(a-b) = \cF(b)$. The $\delta$-function is defined
on conjugacy classes, {\it i.e.} it vanishes unless $g$ and $h$ belong
to the same conjugacy class.

\subsection{The modular $S$ and $T$-matrix}

The characters that appear in the partition functions, i.e. the characters
of the chiral algebra $\cA$, form a representation of the modular group,
generated by $T:\tau \rightarrow \tau+1$
and $S:\tau \rightarrow -1/\tau$:
\begin{eqnarray}
  \label{eq:12}
  \chi_{\L}(\tau+1)= \sum_{M} T_{\L}^{~M}\chi_M(\tau) ~,\\
\chi_{\L}(-1/\tau) = \sum_{M} S_{\L}^{~M} \chi_M(\tau)~.
\end{eqnarray}
The $T$ matrix is diagonal and equals
\bea
    T_{L,M} & = &  \exp \left[
      2\pi\i\left(h_{{\Lambda}} - \frac{c}{24}\right) \right] \d_{\L,M}\\
    & = & \exp \left[2\pi\i
 \left( \frac{(\L,\L + 2 \rho)}{2h} -
   \frac{1}{24}\frac{k d}{h}  \right ) \right] \d_{\L,M}.
\eea
The modular $S$-matrix is unitary and symmetric and equals
\beq
    S_{\L,M} = \cN_\cS \sum_{w\in W} {\rm sign} (w)
    \exp\left[ \frac{-2\pi\i}{h}
      \left(w(\L+\rho),M+\rho\right) \right]~,
\eeq
with normalization
\beq \label{eq:hjs}
    \cN_\cS = e^{2\pi\i \left(\frac{d-r}{8}\right)
    } \left|{L_w \over L^\vee}\right|^{-1/2} \frac{1}{\sqrt{h^r}}~.
\eeq
Here $|L_w/L^{\vee}|$ is the number of weights inside the unit cell of the
coroot lattice.

Comparison with the Weyl character formula eq. (\ref{weylch}) shows that
the modular $S$-matrix is proportional to a character of the horizontal
sub-algebra
\bea
\nonumber
S_{L,M} &=& \cN_S \left( \sum_{w \in W} {\rm sign}(w) \exp \left[ \frac{-2\pi
\i}{h} (w(\rho), M+\rho)\right]\right) \cX_\L\left(-\frac{
(M+\rho,H)}{h} \right) \\
&=& S_{0,M+\rho} \cX_\L \left(-\frac{
(M+\rho,H)}{h} \right)~.
\eea
Note that if we express $M+\rho$ in Dynkin basis and $H$ in Chevalley basis,
then $(M+\rho,H)\equiv \sum_i (M+\rho)_i C^{-1}_{ij} H^j $.

\end{document}